\documentclass[10pt,journal,compsoc]{IEEEtran}

\ifCLASSOPTIONcompsoc
  \usepackage[nocompress]{cite}
\else
  \usepackage{cite}
\fi
\usepackage[final]{graphicx}

\ifCLASSINFOpdf
\else
\fi

\usepackage{algorithm}
\usepackage{algpseudocode}
\newcommand{\setalglineno}[1]{%
  \setcounter{ALG@line}{\numexpr#1-1}}

\usepackage{xspace}
\usepackage{color}
\usepackage{amsmath}
\usepackage{amsfonts} 
\usepackage{amssymb} 
\usepackage{upgreek} 
\usepackage{hyperref}
\usepackage{url}
\usepackage{array}
\usepackage{multirow}
\usepackage{booktabs}

\usepackage{pgfplots} 

\usepackage[mathscr]{euscript}

\usepackage{graphicx}
\usepackage{caption}
\usepackage{subcaption}
\usepackage{eurosym}
\usepackage{balance} 

\newcommand{\mm}[1]{{\small{\textsf{#1}}}}
\newcommand{\etal}[0]{\emph{et al.}\xspace}
\newcommand\ie{\emph{i.e.},\xspace}
\newcommand\eg{\emph{e.g.},\xspace}
\newcommand{\citesec}[1]{Section~\ref{sec:#1}}
\newcommand{\citefig}[1]{Figure~\ref{fig:#1}}
\newcommand{\citealgo}[1]{Algorithm~\ref{algo:#1}}
\newcommand{\citetable}[1]{Table~\ref{table:#1}}

\begin{document}
%
\title{Feature-Model-Guided Online Learning for Self-Adaptive Systems}
%
%
%
%

\author{Andreas~Metzger,
   Cl{\'{e}}ment~Quinton,
	 Zolt{\'{a}}n {\'{A}}d{\'{a}}m Mann, 
	 Luciano~Baresi, and
	 Klaus~Pohl 


\IEEEcompsocitemizethanks{\IEEEcompsocthanksitem A. Metzger, Z. A. Mann, and K. Pohl are with 
paluno (The Ruhr Institute for Software Technology), University of Duisburg-Essen, Essen, Germany.
E-mail: \{andreas.metzger, zoltan.mann, klaus.pohl\}@paluno.uni-due.de
\IEEEcompsocthanksitem C. Quinton is with University of Lille, Lille, France.
E-mail: clement.quinton@univ-lille.fr
\IEEEcompsocthanksitem L. Baresi is with Politecnico di Milano, Dipartimento di Elettronica, Informazione e Bioingegneria, Milan, Italy.
E-mail: luciano.baresi@polimi.it}
\thanks{\copyright~2019 IEEE.  Personal use of this material is permitted.  Permission from IEEE must be obtained for all other uses, in any current or future media, including reprinting/republishing this material for advertising or promotional purposes, creating new collective works, for resale or redistribution to servers or lists, or reuse of any copyrighted component of this work in other works.}
}

%
%

\markboth{~}
{ }
%




\IEEEtitleabstractindextext{%
\begin{abstract}


A self-adaptive system can modify its own structure and behavior at runtime based on its perception of the environment, of itself and of its requirements.
To develop a self-adaptive system, software developers codify knowledge about the system and its environment, as well as how adaptation actions impact on the system.
However, the codified knowledge may be insufficient due to design time uncertainty, and thus a self-adaptive system may execute adaptation actions that do not have the desired effect.
Online learning is an emerging approach to address design time uncertainty by employing machine learning at runtime.
Online learning accumulates knowledge at runtime by, for instance, exploring not-yet executed adaptation actions.
We address two specific problems with respect to online learning for self-adaptive systems.
First, the number of possible adaptation actions can be very large.
Existing online learning techniques randomly explore the possible adaptation actions, but this can lead to slow convergence of the learning process.
Second, the possible adaptation actions can change as a result of system evolution.
Existing online learning techniques are unaware of these changes and thus do not explore new adaptation actions, but explore adaptation actions that are no longer valid.
We propose using feature models to give structure to the set of adaptation actions and thereby guide the exploration process during online learning.
Experimental results involving four real-world systems suggest that considering the hierarchical structure of feature models may speed up convergence by 7.2\% on average.
Considering the differences between feature models before and after an evolution step may speed up convergence by 64.6\% on average.
For a cloud management system, experimental results suggest that this faster convergence leads to energy savings of 12.0\% on average and a reduction in virtual machine migrations by 74.3\% on average.

\end{abstract}

}

\maketitle

\IEEEdisplaynontitleabstractindextext

%
\IEEEpeerreviewmaketitle

\IEEEraisesectionheading{\section{Introduction}\label{sec:introduction}}

A \emph{self-adaptive} system is capable of modifying its own structure and behavior at runtime based on its perception of the environment, of itself and of its requirements~\cite{SESAS_II,SalehieT09,BaresiNG06}.
As an example, take a self-adaptive web application.
Faced with  a sudden increase in workload, the web application may deactivate its resource-intensive recommender engine in order to maintain its performance requirements~\cite{KleinMAH14}. 

\begin{figure}[htbp]
	\centering
		\includegraphics[width=.6\columnwidth]{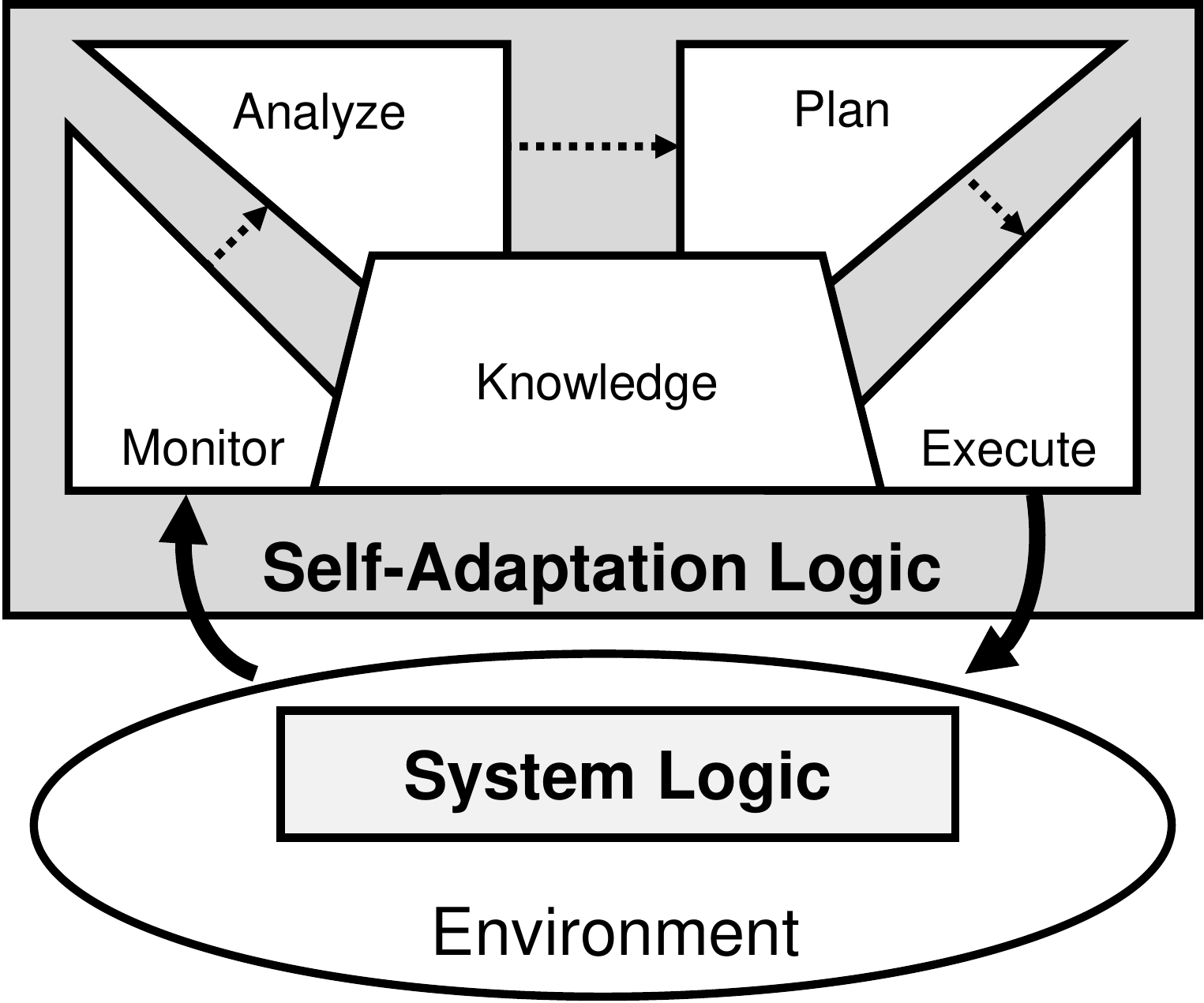}
	\caption{Self-adaptive system reference model (based on~\cite{KephartC03})}
	\label{fig:MAPE}
\vspace{-1em}
\end{figure}

As depicted in Figure~\ref{fig:MAPE}, a self-adaptive system can conceptually be structured into two main elements~\cite{KephartC03,SalehieT09}: the \emph{system logic} (aka. the managed element) and the \emph{self-adaptation logic} (aka. the autonomic manager).
The self-adaptation logic can be further structured into four main conceptual activities that leverage a common \emph{knowledge} base~\cite{IglesiaW15}.
The four activities \emph{monitor} the system and its environment, \emph{analyze} monitored data to determine adaptation needs, \emph{plan} adaptation actions, and \emph{execute} these adaptation actions at runtime.

To populate the self-adaptation logic's knowledge base, software developers codify knowledge about the system and its environment, as well as how adaptation actions impact on the system~\cite{ChenB17,JamshidiPM16,DIppolitoBKMSU14,EsfahaniEM13}.
However, codifying this knowledge at design time may not be fully possible due to design time \emph{uncertainty}~\cite{Esfahani_ESEC_2011}.
For example, when following a \emph{model-based} adaptation approach, 
software developers define analytical models about the system and its environment from which adaptation actions are generated at runtime~\cite{MorenoCGS18,TajalliGEM10,FlochHSELG06}.
However, such analytical models may not be accurate due to simplifying assumptions made at design time~\cite{JamshidiEtAl2019,EsfahaniEM13,DIppolitoBKMSU14}.
As another example, when following a \emph{rule-based} adaptation approach, software developers have to specify adaptation rules prescribing which adaptation action is executed in a given environment situation~\cite{FrommgenRLB15,LaneseBM10,GarlanCHSS04}.
This requires anticipating at design time the potential environment situations the system may encounter at runtime.
However, for many application domains and in particular for open-world systems~\cite{BaresiNG06}, anticipating all potential environment situations at design time is often infeasible~\cite{RamirezJC12}. 

Due to design time uncertainty, insufficient knowledge about the system and its environment may be codified at design time.
As a result, a self-adaptive system may execute adaptation actions that do not have the desired effect, \ie are \emph{ineffective}.
An ineffective adaptation action may have no effect at all, may only have a partial or sub-optimal effect, or may even have a negative effect on the system~\cite{FredericksDC14}.

\begin{figure}[htbp]
	\centering
		\includegraphics[width=.65\columnwidth]{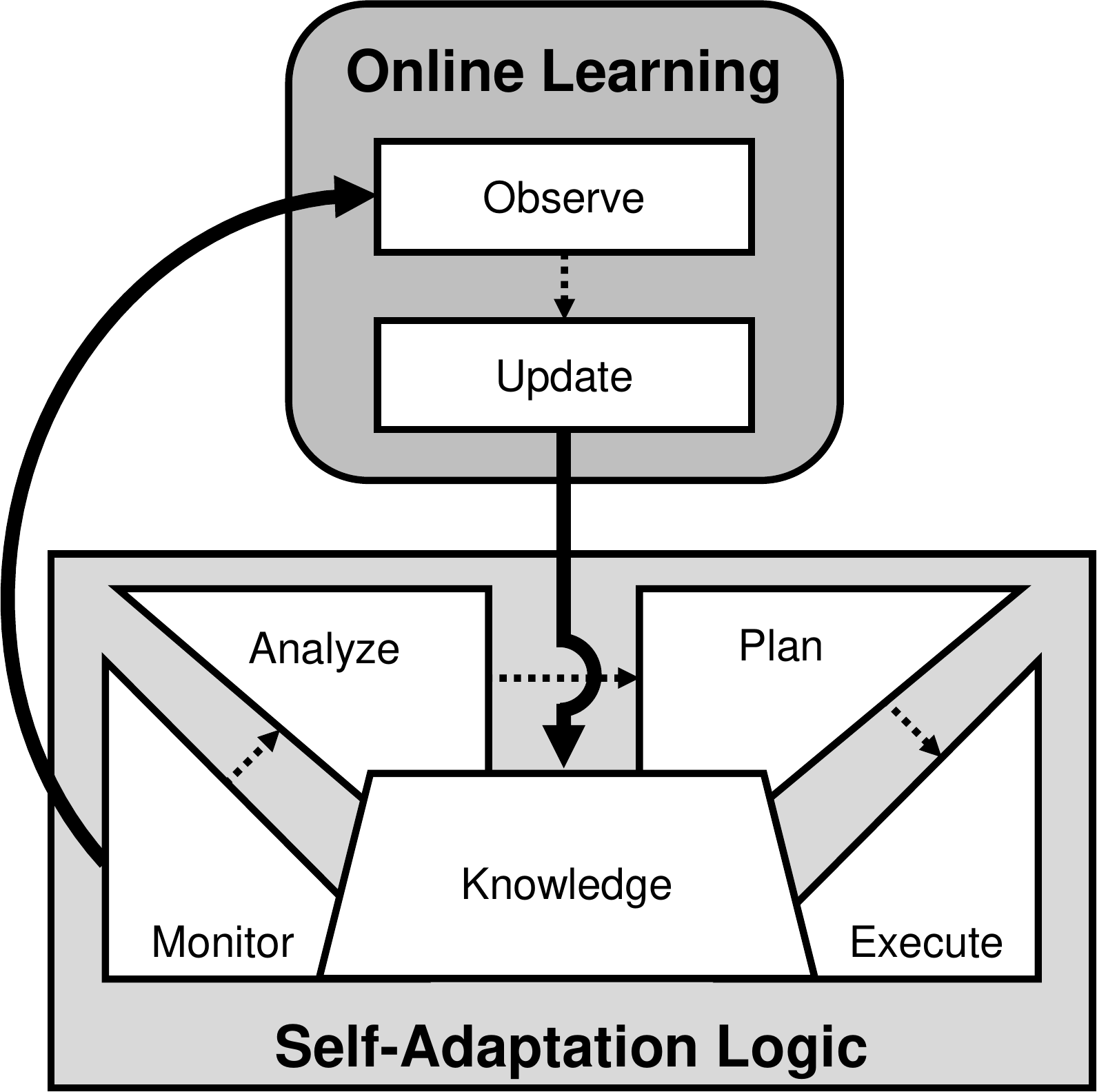}
	\caption{Online learning for self-adaptive systems}
	\label{fig:ol-MAPE}
\end{figure}

\emph{Online learning} is an emerging approach to address design time uncertainty by employing machine learning at runtime.
As depicted in Figure~\ref{fig:ol-MAPE}, online learning \emph{observes} the live system and its environment in order to accumulate knowledge at runtime to \emph{update} the self-adaptation logic's knowledge base.

Online learning was employed for model-based adaptation where the knowledge base includes analytical models of the system and its environment~\cite{EsfahaniEM13,SykesCMKRI13}, as well as for rule-based adaptation where the knowledge base contains adaptation rules~\cite{ZhaoZZJ17,FilhoP17,Qian2015}.
To concisely describe the problem addressed and our contributions, we focus on rule-based adaptation in the remainder of this paper.

\subsection{Problem Statement}
\label{sec:Problem}

The performance of machine learning depends -- to a large degree -- on the amount of data available for learning~\cite{Domingos12,LeCunBH15}.
When machine learning is used for online learning, this data is collected at runtime in order to be representative of the running system and its environment.
Existing findings indicate that when online learning is used to learn adaptation rules, it  typically takes quite many observe-and-update iterations until the learning process converges to a set of effective adaptation rules~\cite{Tesauro07,MoustafaZ14}.\footnote{An effective adaptation rule triggers an effective adaptation action in a given environment situation.}
As an example, when using supervised learning, the system has to collect a sufficient amount of training data that is representative of the system's environment in order to determine possible environment situations.
As another example, when using reinforcement learning, the system has to perform many interactions with its environment in a trial-and-error fashion to explore which adaptation action should be executed in which environment situation.

Until online learning has converged, the system most likely executes inefficient adaptation rules, because not enough observations have yet been made.
In addition to adaptation rules having no effect or being sub-optimal, some of them may lead to negative effects.
As an example, an adaptation action may activate all optional system features, thereby leading to a surge in resource consumption and a degradation of system performance.
Executing these adaption actions has real consequences as they happen in the live system~\cite{FilhoP17}.
If online learning requires a high number of iterations to converge, the impact and costs of online learning can become prohibitive~\cite{Tesauro07}.
How fast online learning converges is thus a very important factor~\cite{KaelblingLM96}.

Different strategies were used to speed up the convergence of online learning for self-adaptive systems.
These strategies include choosing the best-performing variant of a learning algorithm~\cite{amoui2008adaptive}, controlling the rate of how many not-yet-executed adaptation actions are explored~\cite{JamshidiEtAl2016}, employing transfer learning~\cite{Jamshidi_SEAMS_2017}, and using an initial offline learning phase~\cite{TesauroJDB07}.
However, these strategies do not explicitly consider the following two specific properties of a self-adaptive system's \emph{adaptation space} (aka. the set of all possible adaptation actions~\cite{MirandolaPS14,AlferezPMSD14,WeynsI0M18}):

\textbf{Large adaptation space.} To update the knowledge base, existing online learning approaches for rule-based adaptation randomly explore the adaptation space by selecting not-yet executed adaptation actions.\footnote{See the approaches~\cite{amoui2008adaptive,KimP09,dutreilh2011using,BarrettHD13,BuRX13,JamshidiEtAl2016,CaporuscioDGM16,FilhoP17,WangCWYHZB17,Tesauro07,XuRB12,MoustafaZ14,ZhaoZZJ17} described in Section~\ref{sec:related_work}.}
The speed of convergence depends on the size of the adaptation space, because each not-yet executed adaptation action has an equal chance of being selected.
If the adaptation space is small, the speed of convergence of online learning using such random adaptation action selection can be acceptable.
However, the adaptation space of a self-adaptive system can be large~\cite{QuinEtAl2019,JamshidiEtAl2019}. 
In such a case, random adaptation action selection can lead to slow convergence~\cite{RL,BuRX13,FilhoP17}.

There exist machine learning techniques that can cope with a large space of actions.
However, these techniques require the space of actions to be continuous. 
A continuous space of actions is represented by continuous variables, such as real-valued variables.
Setting a specific angle for a robot arm or changing the set-point of a thermostat are examples for a continuous space of actions~\cite{KaelblingLM96}.
Many kinds of self-adaptive systems have a non-continuous space of adaptation actions.
Examples include architecture-based self-adaptive systems, where adaptation actions are changes of component compositions~\cite{GarlanCHSS04}, and dynamic software product lines, where adaptation actions are the activation and deactivation of system features~\cite{DSPL-Ghezzi}.
As an example, take a system that offers ten optional features that may be dynamically activated and deactivated in any combination and that allows changing from any active feature combination to any other possible feature combination.\footnote{We assume that there are no technical or logical constraints on the order of adaptation actions. We discuss this further in Section~\ref{sec:Limitations}.}
Its adaptation space thus contains $2^{10} = 1024$ adaptation actions.
These $1024$ adaptation actions cannot be represented as a continuous variable.


\textbf{Change of adaptation space due to system evolution.} 
Existing online learning approaches for rule-based adaptation are unaware of system evolution~\cite{KinneerCWGG18,SEAMS16}.
They do not consider the fact that a self-adaptive system -- like any software system -- may undergo evolution~\cite{EvolvingDSPL-SPLC15,Gilles12}.
Self-adaptation refers to the automatic modification of the system by itself.
Evolution refers to the modification of the system by humans~\cite{Ghezzi17,IGI_Agile_Book}.
During evolution, software developers may modify the system to correct bugs, remove rarely used features, or introduce new features~\cite{Bosch-SEAMS16}.
System evolution means that the adaptation space may change.
In the example from above, one of the ten features may be removed in an evolution step, thereby reducing the system's adaptation space.
As another example, a new feature may be introduced in an evolution step, thereby adding new possible feature combinations and thus adaptation actions to the adaptation space.

There exist machine learning techniques that can cope with environments that change over time (so called non-stationary environments~\cite{SugiyamaK12}).
However, they do not consider that the space of actions may change over time.
Being unaware of the changes to the adaptation space  means that these techniques may explore adaptation actions that are no longer valid and thus may even have negative effects on the system.
Also, they are unaware of the new adaptation actions and thus will not select these new adaptation actions even though they may lead to effective adaptation rules.
A simple solution would be to restart the online learning process from scratch after each evolution step.
However, this means that knowledge already gained is lost, and thus cannot be used to speed up the convergence of online learning after an evolution step.

\subsection{Contributions}
\label{sec:contrib}

We introduce \emph{online learning strategies} that address potentially large adaptation spaces and that can cope with a change of the adaptation space due to system evolution.
Our online learning strategies use \emph{feature models}~\cite{variability2013,SPL-FOSE14,EsfahaniEM13,HincheyPS12} to give structure to the system's adaptation space.
A feature model is a tree or a directed acyclic graph of features, organized hierarchically.
Feature models thereby provide additional information to guide the online learning process.
Concretely, we make the following two contributions.

\textbf{Using the feature model structure to explore the adaptation space.} 
Our main idea is to leverage the hierarchical structure of the feature model to take more informed decisions during the exploration of the adaptation space than random exploration does.
The strategies we propose systematically traverse the feature model to select the next adaptation action to be executed and observed.
To illustrate, our strategies may first explore the different sub-features of a parent feature, before exploring features which are not directly related to the parent feature.
We argue that via such systematic exploration, our learning strategies can speed up convergence.
	
\textbf{Using feature model deltas to capture changes in the adaptation space due to system evolution.} 
Our main idea is to make online learning aware of changes in the adaptation space by using feature model deltas.
The strategies we propose analyze the delta between a feature model before and after an evolution step.
Thereby, our strategies can identify  added and removed adaptation actions.
Removed adaptation actions are no longer considered, while added adaptation actions are targeted first, as they may offer new opportunities for finding effective adaptation rules.
In addition, our strategies systematically reuse past knowledge about whether the presence of a certain feature contributed to an effective adaptation rule.
We argue that by considering new adaptation actions and by reusing past knowledge, our learning strategies can speed up convergence.

We experimentally assess our online learning strategies using four real-world systems.
We compare our experimental results with the results for random exploration of the adaptation space as a baseline. 

The remainder of this paper is organized as follows. 
\citesec{background} explains the key concepts of feature models and rule-based adaptation, and also introduces a running example.
\citesec{random} describes and illustrates the random exploration of the adaptation space to serve as baseline for our contributions and experiments.
\citesec{adapt} explains how we use the feature model structure to systematically explore the adaptation space.
\citesec{evol} explains how we use feature model deltas to capture adaptation space changes due to system evolution.
\citesec{setup} presents the design and results of our experiments.
\citesec{related_work} analyzes related work.
\citesec{conclusion} concludes with a discussion on limitations and directions for future work. 

\section{Fundamentals and Running Example}
\label{sec:background}
This section explains the key concepts of feature models and rule-based adaptation, which are illustrated by a running example of a web application.

\subsection{Feature Models for Self-adaptive Systems}
\label{sec:FM}

As introduced above, our online learning strategies exploit additional knowledge about the self-adaptive system's adaptation space encoded in the form of feature models.

A \emph{feature model} is a tree or a directed acyclic graph of features~\cite{SPL-FOSE14,MetzgerHPSS07}, organized hierarchically. 
A feature can be decomposed into mandatory, optional or alternative sub-features. 
A mandatory sub-feature has to be activated if its parent feature is activated.
An optional sub-feature may or may not be activated if its parent feature is activated.
At least one of the alternative sub-features has to be activated if their parent feature is activated.
Additional constraints, such as ``excludes'' or ``requires'' constraints, between two features, express inter-feature dependencies.
Thereby, a feature model describes the possible and allowed feature combinations of a system.

While feature models are traditionally used in software product line engineering to define the set of systems of the product line at design time~\cite{SPL-FOSE14}, dynamic software product lines extend the use of feature models to runtime~\cite{DSPL-Ghezzi,EsfahaniEM13,HincheyPS12}.
In dynamic software product lines, the feature model describes the set of possible adaptation actions in the form of feature combinations, \ie set of to be activated and deactivated system features.
In a similar way, feature models can be used for architecture-based self-adaptive systems to define the possible runtime compositions of system components~\cite{MorinBJFS09,GomaaH07}.
A feature model thereby can be used to define a self-adaptive system's adaptation space, where each adaptation action is expressed in terms of the feature combination to be active after adaptation.

\citefig{fm} shows the feature model of a self-adaptive web application we use as a running example.
The \mm{DataLogging} feature is mandatory (which means it is always active), while the \mm{ContentDiscovery} feature is optional.
The \mm{DataLogging} feature has three alternative sub-features, which express that at least one of the three levels of data logging must be active: \mm{Min}, \mm{Medium} or \mm{Max}.
The \mm{ContentDiscovery} feature has two optional sub-features \mm{Search} and \mm{Recommendation}.
The constraint \mm{Recommendation} $\Rightarrow$ \mm{Max} $\vee$ \mm{Medium} specifies that a sufficient level of data logging is required to collect enough information about the web application's users and transactions to make good recommendations.

\begin{figure}[!h]
\centering
	\includegraphics[width=0.7\columnwidth]{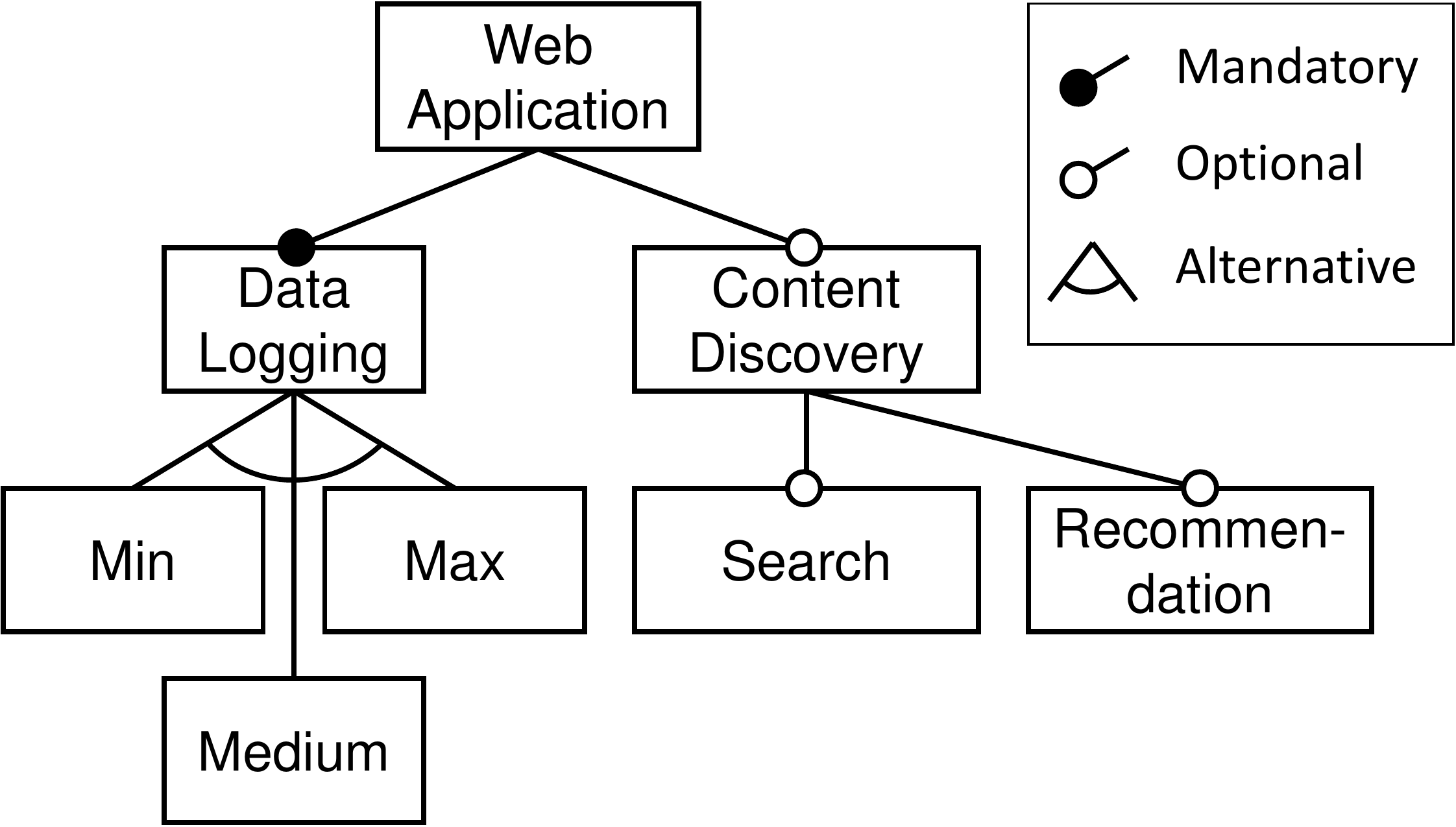}

$\textsf{Recommendation} \Rightarrow \textsf{Max} \vee \textsf{Medium} $

\caption{Feature model of self-adaptive web application.}
\label{fig:fm}
\end{figure}
\vspace{-1em}

\subsection{Rule-based Adaptation}
\label{sec:rule-based-adaptation}

In rule-based adaptation, adaptation rules specify which adaptation action is executed in response to a given environment situation~\cite{FrommgenRLB15,LaneseBM10,GarlanCHSS04}.
To illustrate, let us consider that the web application introduced above has to adapt to changes in its environment in order to maintain its performance requirements.
More concretely, the web application should adapt to changing workloads (\ie number of simultaneous users) in order to keep its response time below 500ms.
A software developer may express an adaptation rule for the web application such that it turns off some of the features in the presence of a higher workload, thereby reducing the resource needs of the application.

\citefig{adaptation} shows a concrete example.
Let us assume a software developer has specified an adaptation rule that states if the system faces and environment situation of more than $1000$ concurrent users, then the \mm{Search} feature should be deactivated.
Here, the software developer estimates that deactivating the \mm{Search} feature will lead to a sufficient reduction in resource needs.
As shown in the figure, if the system runs in feature combination \{\mm{DataLogging}, \mm{Max}, \mm{ContentDiscovery}, \mm{Search}, \mm{Recommendation}\}, this adaptation rule results in adapting the system to feature combination \{\mm{DataLogging}, \mm{Max}, \mm{ContentDiscovery}, \mm{Recommendation}\}.

\begin{figure}[!t]
\centering
\includegraphics[width=\columnwidth]{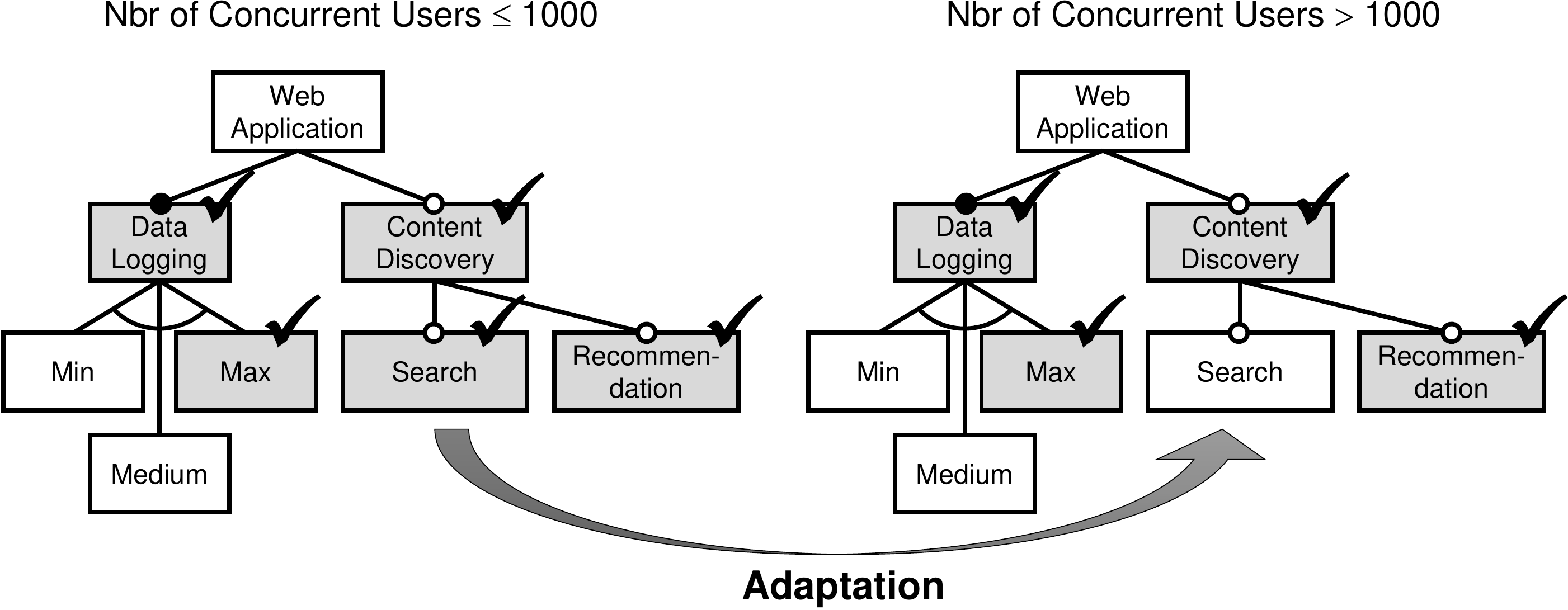}

\caption{An example of feature-based runtime adaptation.}
\label{fig:adaptation}
\vspace{-1em}
\end{figure}

We formally define the \emph{effectiveness} of an adaptation rule using Zave and Jackson's framework applied to self-adaptive systems as presented in~\cite{CalinescuGKM12}. 
Let $E$ be the environment situation that triggers the rule, $S$ the self-adaptive system after the execution of the adaptation action specified in the rule, and $R$ the system's requirements. 
We consider an adaptation rule effective if $S, E \models R$.
This means the rule is effective if the system after adaptation meets its  requirements.

\section{Random Adaptation Space Exploration as Baseline}
\label{sec:random}

As explained in Section~\ref{sec:introduction}, existing online learning strategies for rule-based adaptation randomly explore the adaptation space in order to select not-yet executed adaptation actions.
In this section we thus introduce the random online strategy as a baseline against which we compare our online learning strategies.

\subsection{Illustration}
\label{sec:RandomLearningInRunningExample}

Online learning in our running example from Section~\ref{sec:background} observes whether the adaptation rule to deactivate the \mm{Search} feature in the presence of more than $1000$ users is effective.
This means online learning observes whether the system after executing this adaptation rule is able to meet the response time requirements given the increased number of users.
This adaptation rule may turn out to be ineffective, because 
only turning off the \mm{Search} feature may not be  sufficient to meet the response time requirements.

If the adaptation rule is ineffective, a random online learning strategy explores different alternative adaptation actions at random, until an adaptation action is found that is effective in the given environment situation.
Table~\ref{tab:Random} illustrates such a random exploration of the adaptation space.
We assume that only the feature combination \{\mm{DataLogging}, \mm{Min}, \mm{ContentDiscovery}, \mm{Search}\} is able to meet the response time requirements.
In the example it takes six iterations until this effective feature combination is found and thus will be used as adaptation action in the adaptation rule.

\begin{table}[htbp]
	\centering
					\includegraphics[width=1.00\columnwidth]{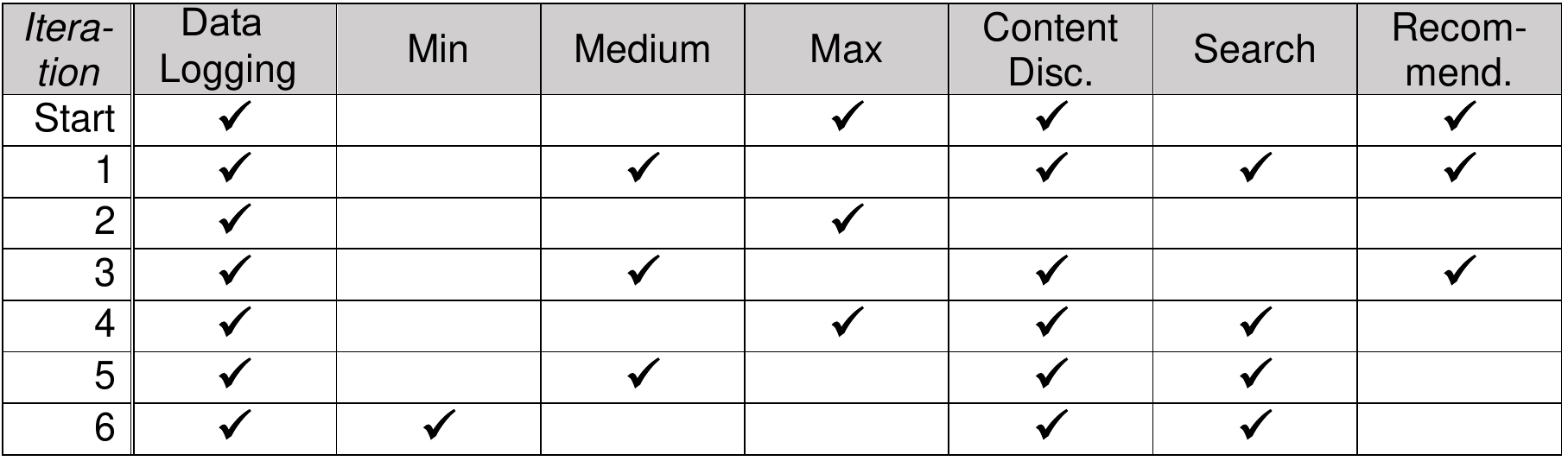}
	\caption{Example of random adaptation space exploration (check mark indicates active feature). }
	\label{tab:Random}
\end{table}
\vspace{-1em}


\subsection{Realization}
\label{sec:rand}

Below we provide a potential realization of a random learning strategy -- we call Rand -- to serve as baseline realization for our experiments in Section~\ref{sec:setup}.
The Rand strategy chooses a new feature combination randomly from the adaptation space of the self-adaptive system.
Given a feature model $\mathcal{M}$ that specifies the adaptation space, with $\mathcal{F}$ being the non-empty set of features of $\mathcal{M}$, the Rand strategy is realized by the iterative \citealgo{random}.

%
%
%
%

\begin{algorithm}
\caption{Random Strategy (Rand)}\label{algo:random}
\begin{algorithmic}[1] 

\State Set$<$Set$<$Feature$>$$>$ $\mathcal{C}_{f}$, $\mathcal{C}'_{f} = \emptyset$;
\State Set$<$Feature$>$ $\mathcal{F'}$, \emph{comb};
\State Feature $f$;\\

\Function{Explore}{FeatureModel $\mathcal{M}$, Set$<$Feature$>$ $\mathcal{F}$}

	\State $\mathcal{F'} \leftarrow \mathcal{F}$;

	\Repeat 
	\State $f\gets $ randomSelect$(\mathcal{F'})$; \label{line:1A}
	\State $\mathcal{C}_{f} \gets$ getFeatureCombinations($\mathcal{M}$, $f$) $\setminus$ $\mathcal{C}'_{f}$;\label{line:1B}
	\If {$\mathcal{C}_{f} \neq \emptyset$} 
		\State \emph{comb} $\gets $ randomSelect($\mathcal{C}_{f}$);\label{line:1C}
		 \If {\emph{effective}(\emph{comb})} 
		\State {\textbf{return}};
		 \EndIf
		\State $\mathcal{C}'_{f} \gets \mathcal{C}'_{f}$ $\cup$ \{\emph{comb}\};\label{line:1X}
	\Else \State $\mathcal{F'} \leftarrow \mathcal{F'} \setminus \{f\}$;\label{line:1D}
	\EndIf
	\Until $\mathcal{F'} = \emptyset$;
\EndFunction
\end{algorithmic}
\end{algorithm}

Note that the realization of the Rand strategy only uses the feature model to address the problem that enumerating all feature combinations of a large adaptation space may not scale due to its exponential complexity~\cite{BenavidesTC05}.
It does not use the structure of the feature model to systematically explore the adaptation space.

The algorithm first selects a feature $f$ randomly (line~\ref{line:1A}), \eg the feature~\mm{Recommendation} in our running example from Section~\ref{sec:background}.
Then the algorithm determines all possible feature combinations $\mathcal{C}_{f}$ containing $f$ (line~\ref{line:1B}) that were not previously selected (line~\ref{line:1B} and line~\ref{line:1X}). 
For instance, if \mm{Recommendation} is selected, then \mm{Max} or \mm{Medium} must be selected and \mm{Search} can possibly be selected as well (but not \mm{Min} because of the constraint expressed in the feature model; see Figure~\ref{fig:fm}). 
To realize the \emph{getFeatureCombinationWith()} operation, we rely on the observation that computing all possible feature combinations beginning with a partial feature combination may scale better than computing all possible feature combinations from scratch~\cite{Benavides_2010}.

If several possible feature combinations exist (\ie if $|\mathcal{C}_{f}| > 1$), one feature combination is selected randomly (line~\ref{line:1C}). 
If no effective feature combination among $\mathcal{C}_{f}$ is found, the strategy starts over by selecting another feature and continues as long as no other new feature is available. 

At the end of each iteration the selected feature $f$ is removed from the set of all features (line~\ref{line:1D}).
Together with not visiting again an already visited feature combination (line~\ref{line:1B} and line~\ref{line:1X}), this effectively implements a random selection \emph{without }replacement.
Such a random selection without replacement is important to ensure a fair baseline for the comparisons in Section~\ref{sec:setup}, as our strategies also select without replacement.

\section{Using the Feature Model Structure to Explore the Adaptation Space}
\label{sec:adapt}

In this section we explain and illustrate our online learning strategies that use the feature model structure to systematically explore the adaptation space.

\subsection{Solution Idea and Illustration} 
\label{sec:idea-size}

As introduced above, the main solution idea for our online learning strategies is to systematically explore the adaptation space by hierarchically traversing the feature model.
Below, we introduce two variants of online learning strategies that differ in the way the feature model is traversed.

\textbf{Incremental Strategy (Inc).}
This strategy takes advantage of the semantics typically encoded in the structure of feature models.
Non-leaf features in a feature model are usually abstract features used to better structure variability~\cite{Thum_SPLC_11}. 
These abstract features often do not have an impact at implementation level, but delegate their realization to their sub-features. 
Sub-features thus may offer different realizations of their abstract parent feature.
The sub-features of a common parent feature, \ie \emph{sibling} features, can thus be considered semantically connected.
In our running web application example (see \citefig{fm}), the \mm{ContentDiscovery} feature has two sub-features \mm{Search} and \mm{Recommendation} offering different concrete ways how a user may discover online content.
The idea behind the Inc strategy is to exploit the information about these potentially semantically connected sibling features and systematically explore 
them first before exploring other features.

To illustrate the Inc strategy, let us start with feature combination \{\mm{DataLogging}, \mm{Max}, \mm{ContentDiscovery}, \mm{Recommendation}\} of the ineffective adaptation rule from Section~\ref{sec:random}.
The Inc strategy first explores sibling features starting from this feature combination.
In our example, let us say the Inc strategy starts exploration of the two sibling features \mm{Recommendation} and \mm{Search}.\footnote{Note that this entails some random selection of whether to start with the sub-features of \mm{\footnotesize{DataLogging}} or \mm{\footnotesize{ContentDiscovery}}.}
The Inc strategy systematically explores all feature combinations involving the \mm{Recommendation} feature, and then moves to systematically exploring all feature combinations involving the \mm{Search} feature.
Table~\ref{tab:Inc} shows a typical exploration sequence of adaptation actions of the Inc strategy (with the step-wise exploration of sibling features highlighted in gray).
In this case, it takes 5 iterations until an effective adaptation action is found (one less than in the random search example from Section~\ref{sec:random}).

\begin{table}[htbp]
		\centering
		\includegraphics[width=1.00\columnwidth]{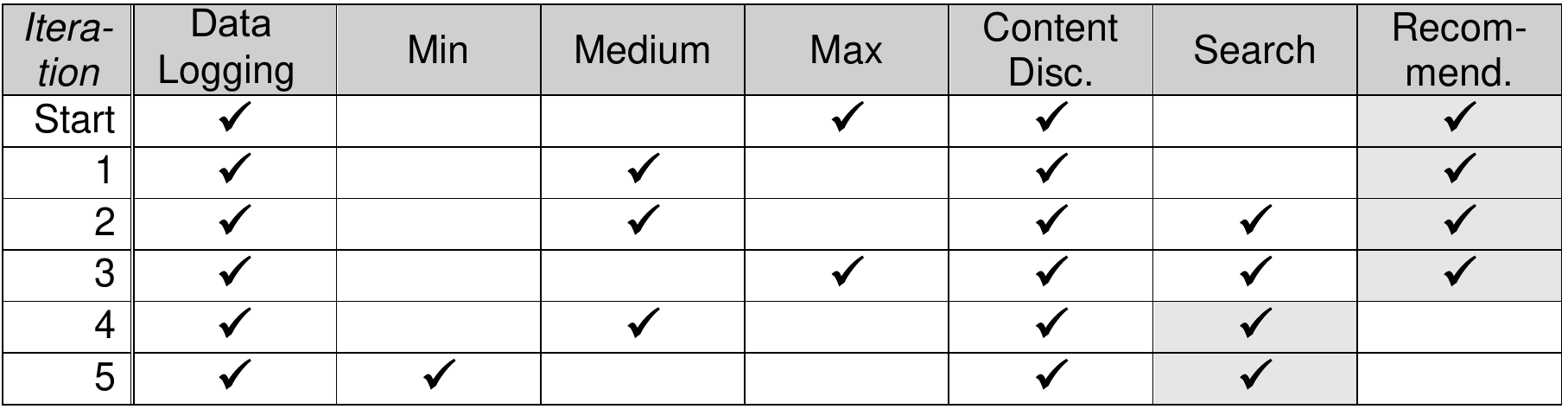}
	\caption{Example of adaptation space exploration via the incremental strategy (Inc). }
	\label{tab:Inc}
\end{table}

\textbf{Feature Degree Strategy (Deg).} Even though the Inc strategy makes use of the structure and hierarchy of the feature model, it still contains several random elements.
In particular, it randomly determines the order in which sibling features are explored.
To take a more informed decision about which of the sibling features to explore, we define the Deg strategy which makes use of the concept of \emph{feature degree}.
We define the feature degree for a given feature $f$ as the number of feature combinations that contain $f$. 
The intuition here is that there may be a higher probability of finding an effective feature combination when considering features with high feature degrees, as they are present in more feature combinations. 

In our example, the feature degree of the \mm{Search} feature  is 5, while of the \mm{Recommendation} feature it is only 4 (due to the constraint requiring at least the \mm{Medium} logging level).
The Deg strategy thus first explores all feature combinations involving the \mm{Search} feature before exploring other feature combinations.
Table~\ref{tab:Deg} shows a typical exploration sequence of the Deg strategy (with the exploration of the sibling feature with the highest feature degree highlighted in gray).
In this case, it takes 4 iterations until the effective adaptation action is found (one less than for the Inc strategy).

\begin{table}[h]
		\centering
		\includegraphics[width=1.00\columnwidth]{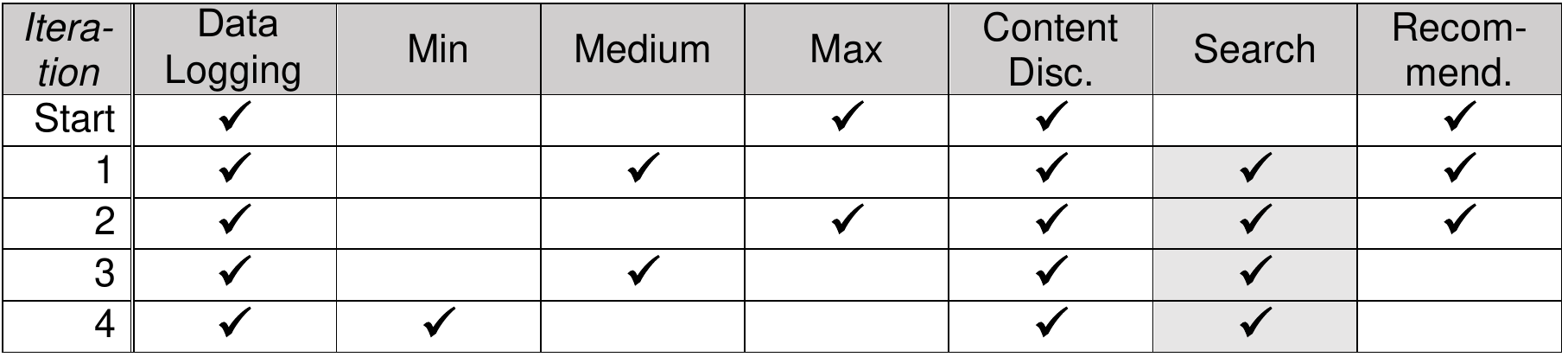}
	\caption{Example of adaptation space exploration via the feature degree strategy (Deg). }
	\label{tab:Deg}
\end{table}

Note that the above examples were purposefully chosen to show the potential improvements of our strategies.
As we will experimentally analyze and discuss in Section~\ref{sec:setup}, there may be some situations in which our strategies may not speed up convergence.

\subsection{Realization}
\label{sec:evol-strategy}

In what follows we explain how we realize the above learning strategies.

\textbf{Incremental Strategy (Inc).}
The Inc strategy  is realized by the recursive \citealgo{incremental}.
The algorithm is initialized by randomly selecting an arbitrary leaf feature $f$ (\ie a feature with no sub-features) among all leaf features that are part of the current feature combination (lines~\ref{line:2A}--\ref{line:2Ax}).\footnote{
The \emph{getLeaves()} function computes the leaf features for a feature model which only includes the active features of \emph{currentFeatureCombination}.
Thereby, finding a leaf feature is always possible.}
Then, the set of feature combinations $\mathcal{C}_{f}$ containing feature $f$  is computed (line~\ref{line:2B}), while the sibling features of feature $f$ are gathered into a dedicated \emph{siblings} set (line~\ref{line:2C}). 

\begin{algorithm}
\caption{Incremental Strategy (Inc)}\label{algo:incremental}
\begin{algorithmic}[1] 
\State Set$<$Feature$>$ \emph{leaves}, \emph{comb};
\State Feature $f$;\\

\Function{Explore}{FeatureModel $\mathcal{M}$, Set$<$Feature$>$ \emph{currentFeatureCombination}}

\State \emph{leaves } $\gets $ getLeaves$(\emph{currentFeatureCombination})$;\label{line:2A}
\State $f\gets $ randomSelect(\emph{leaves});\label{line:2Ax}
\State $\mathcal{C}_{f} \gets$ getFeatureCombinations($\mathcal{M}$, $f$);\label{line:2B}
\State \emph{siblings} $\gets$ siblings($f$);\label{line:2C}
\\
	 \State \textsc{ExploreRecursively}($\mathcal{M}$, $\mathcal{C}_{f}$, \emph{siblings});

\EndFunction
\\
\Function{ExploreRecursively}{FeatureModel $\mathcal{M}$, Set$<$Set$<$Feature$>$$>$ $\mathcal{C}_{f}$, Set$<$Feature$>$ \emph{siblings}}

\If{$\mathcal{C}_{f} \neq \varnothing$} \label{line:2D}
	\State \emph{comb} $\gets $ randomSelect($\mathcal{C}_{f}$);
  \State $\mathcal{C}_{f} \gets \mathcal{C}_{f} ~\setminus$ \{\emph{comb}\};
			 \If {\emph{effective}(\emph{comb})} 
		\State {\textbf{return}};
	\Else 
	 	 \State \textsc{ExploreRecursively}($\mathcal{M}$, $\mathcal{C}_{f}$, \emph{siblings});

		\EndIf \label{line:2E}

\Else 
	\If{\emph{siblings} $\neq \varnothing$} \label{line:2F}
   	\State $f$ $\gets $ randomSelect(\emph{siblings});\label{line:2X}
	  \State \emph{siblings} $\gets $ \emph{siblings} $\setminus$ \{$f$\};
			
		\State $\mathcal{C}_{f} \gets$ getFeatureCombinations($\mathcal{M}$, $f$);
	 \State \textsc{ExploreRecursively}($\mathcal{M}$, $\mathcal{C}_{f}$, \emph{siblings});
 \label{line:2G}
	\Else 
		\If{\emph{parent}($f$) $\neq \emptyset$}\label{line:2H}
			\State $f \gets$ parent($f$);
			\State \emph{siblings} $\gets$ siblings($f$);
			\State $\mathcal{C}_{f} \gets$ getFeatureCombinations($\mathcal{M}$, $f$);
	 \State \textsc{ExploreRecursively}($\mathcal{M}$, $\mathcal{C}_{f}$, \emph{siblings});

		\Else 
		  	\State // Root feature reached
				\State {\textbf{return}};

			\EndIf\label{line:2I}
		\EndIf
	\EndIf

\EndFunction
\end{algorithmic}
\end{algorithm}

While $\mathcal{C}_{f}$ is non-empty, the strategy explores one randomly selected feature combination from $\mathcal{C}_{f}$ and removes the selected feature combination from $\mathcal{C}_{f}$ (lines~\ref{line:2D}--\ref{line:2E}).
If $\mathcal{C}_{f}$ is empty, then a new set of feature combinations containing a sibling feature of $f$ is randomly explored, provided such sibling feature exists (lines~\ref{line:2F}--\ref{line:2G}).
If no feature combination containing $f$ or a sibling feature of $f$ is found, then the strategy moves on to the parent feature of $f$.
Moving to a respective parent feature is repeated until the root feature is reached (lines~\ref{line:2H}--\ref{line:2I}). 

\textbf{Feature Degree Strategy (Deg).}
The Deg strategy is realized by modifying \citealgo{incremental} to make use of the feature degree as shown in \citealgo{deg}.
On the one hand, the feature degree is used to determine which leaf feature to start the learning from.
Instead of randomly selecting a leaf feature, as done in the Inc algorithm (lines~\ref{line:3A}--\ref{line:3B}), the Deg strategy selects a leaf feature with the highest feature degree.
On the other hand, instead of randomly choosing sibling features as done in the Inc algorithm (line~\ref{line:2X}), the Deg strategy uses the feature degree to define an order in which sibling features are chosen (starting with the sibling feature with the highest feature degree). 

To realize the \emph{confDeg} operation, off-the-shelf feature model analysis tools (\eg see~\cite{Benavides_2010}) can be used to compute the number of possible feature combinations containing $f$. 

\begin{algorithm}
\caption{Feature Degree Strategy (Deg)}\label{algo:deg}
\begin{algorithmic}[1] 
\Statex [...]
\setalglineno{5} \State \emph{leaves } $\gets $ getLeaves$(\emph{currentFeatureCombination})$;\label{line:3A}
\State $f\gets $ randomSelect($\{f : g \in leaves \Rightarrow \textrm{confDeg}(f) \geq \textrm{confDeg}(g)\}$);\label{line:3B}
\Statex [...]
%
\setalglineno{23} \State \hspace{3em} \textbf{if} \emph{siblings} $\neq \varnothing$ \textbf{then} \label{line:3F}
   	\State \hspace{4.5em} $f$ $\gets $ randomSelect($\{f : g \in siblings \Rightarrow $
		\Statex \hspace{4.5em} $\textrm{confDeg}(f) \geq \textrm{confDeg}(g)\}$);\label{line:3X}
\Statex [...]
\end{algorithmic}
\end{algorithm}

\section{Using Feature Model Deltas to Capture Adaptation Space Changes due to Evolution} 	
\label{sec:evol}

In this section we explain and illustrate our online learning strategies that use the feature model deltas to capture adaptation space changes due to system evolution.

\subsection{Solution Idea and Illustration} 
\label{sec:idea-evol}

As introduced above, our main solution idea to capture a change in the adaptation space due to system evolution is to use feature model deltas.
We do so by analyzing the delta between a feature model before and after an evolution step.
Thereby we can identify new possible adaptation actions that were added to the adaptation space, as well as adaptation actions that were removed from the adaptation space.
Removed adaptation actions are no longer explored, while added adaptation actions are targeted first, as they may offer new opportunities for finding effective adaptation rules.

Let us assume the adaptation space changes from an adaptation space $\mathcal{A}$ before an evolution step to an adaptation space $\mathcal{A'}$ after an evolution step. 
Given a feature model $\mathcal{M}$ that specifies $\mathcal{A}$ (\ie the set of all possible feature combinations) and a feature model $\mathcal{M'}$ that specifies $\mathcal{A'}$, then two main types of changes of the adaptation space can be detected as deltas between feature models $\mathcal{M}$   and $\mathcal{M'}$.\footnote{A modification of a feature's implementation is not visible in a feature model. 
We discuss this further in Section~\ref{sec:Limitations}.}

\emph{Added feature combinations.} 
	New features may be added to $\mathcal{M'}$ or existing constraints may be removed or relaxed from $\mathcal{M}$ (such as ``requires'' or ``excludes'' constraints).
	This means that new feature combinations are added to the adaptation space $\mathcal{A'}$.
	As an example in our web application, a new sub-feature \mm{Optimized} might be added to the \mm{DataLogging} feature, providing a more resource efficient logging implementation.
 Thereby, new feature combinations are added to the adaptation space, such as \{\mm{DataLogging}, \mm{Optimized}, \mm{ContentDiscovery}, \mm{Search}\}.
As another example, the \mm{Recommendation} implementation may have been improved and it now can work with the \mm{Min} logging feature.
This relaxes the initial constraint as shown in Figure~\ref{fig:fm}, and adds new feature combinations such as \{\mm{DataLogging}, \mm{Min}, \mm{ContentDiscovery}, \mm{Recommendation}\}.

 \emph{Removed feature combinations.} 
	Symmetrical to the above, features from $\mathcal{M}$ may be removed or constraints may be added or tightened in $\mathcal{M'}$.
 This means that feature combinations are removed from the adaptation space.
	
The idea of using feature model deltas to capture these adaptation space changes is two-fold. 
On the one hand, our strategies first explore the feature combinations that were added to the adaptation space by an evolution step, and then explore the remaining feature combinations if needed, \ie we first explore feature combinations from $A' \setminus A$. 
The rationale is that added feature combinations might offer new opportunities to find effective adaptation actions and thus should be explored first.

On the other hand, our strategies accumulate knowledge across the evolution steps about whether the presence of a certain feature may help maintain the system's requirements or not.
The strategies first explore feature combinations from $A' \setminus A$ that include as many as possible features that were part of effective feature combinations before an evolution step and as little as possible features that were part of ineffective feature combinations before an evolution step.

Table~\ref{tab:EvoDeg} shows a typical exploration sequence of such an evolution-aware strategy.
Online learning targets the new, more resource-efficient \mm{Optimized} feature, and in addition, chooses a feature combination that does not  include the \mm{Recommendation} feature, as this was not in any feature combination able to meet the response time requirements so far.
In this case, it would thus only take 1 iteration until the effective adaptation action is found.

\begin{table}[htbp]
		\centering
		\includegraphics[width=1.00\columnwidth]{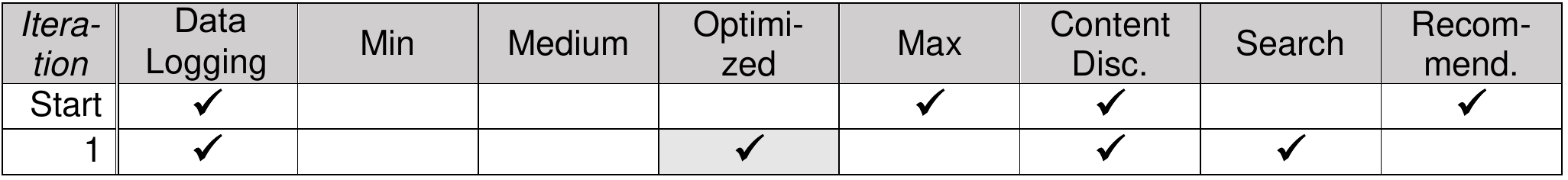}
	\caption{Example of adaptation space exploration via an evolution-aware strategy. }
	\label{tab:EvoDeg}
	\vspace{-2em}

\end{table}

\subsection{Realization}
\label{sec:ExploEvo}

By extending the above three learning strategies Rand, Inc and Deg, we realize three \emph{evolution-aware} learning strategies: EvoRand, EvoInc and EvoDeg.

To exploit the knowledge across evolution steps whether the presence of a certain feature helps maintain the system's   requirements or not, we encode this knowledge in two evolving sets of features:
\begin{itemize}
	\item $\mathcal{F}_{(+)}$ includes features of $\mathcal{M'}$ that were part of at least one effective feature combination, as well as features that were not yet activated in any feature combination.
\item $\mathcal{F}_{(-)}$ includes features of $\mathcal{M'}$ that were only part of ineffective feature combinations. 

\end{itemize}
Based on the actual observations of online learning, these sets of features are updated accordingly. 
Features thus may move from one set to the other over the course of learning and evolution.
In addition, features may be removed or added to the sets due to the deltas between $\mathcal{M}$  and $\mathcal{M'}$.

Our  strategies prioritize feature combinations that include features from $\mathcal{F}_{(+)}$ and do not include any feature from $\mathcal{F}_{(-)}$.
To this end, we extend the algorithms from Section~\ref{sec:evol-strategy} along two directions.
First, we select features from $\mathcal{F}_{(+)}$ to determine the feature combinations to be explored.
Second, we first explore those new feature combinations that do not contain features from $\mathcal{F}_{(-)}$. 

To select features from $\mathcal{F}_{(+)}$, the algorithms are extended as follows.

\textbf{Evolution-aware random strategy (EvoRand).} 
Other than in \citealgo{random} (line~\ref{line:1A}), feature $f$ is first randomly selected among the features in $\mathcal{F}_{(+)}$, and not randomly selected among all possible features in $\mathcal{F}$.

\textbf{Evolution-aware incremental strategy (EvoInc).} 
Other than in \citealgo{incremental} (lines~\ref{line:2Ax} and~\ref{line:2X}), feature $f$ is first randomly selected among those leaf or sibling features that are in $\mathcal{F}_{(+)}$, and not randomly selected among all leaf or sibling features in $\mathcal{F}$.

\textbf{Evolution-aware feature degree strategy (EvoDeg).}
Other than in \citealgo{deg} (lines~\ref{line:2Ax} and~\ref{line:2X}), feature $f$ is first randomly selected among those leaves or sibling features with the highest feature degree that are also in $\mathcal{F}_{(+)}$, and not randomly selected among all leaves or sibling features in $\mathcal{F}$ with the highest feature degree.

To realize first exploring new feature combinations  that do not contain features from $\mathcal{F}_{(-)}$, the algorithms are extended as follows.
Whenever computing the set of feature combinations $\mathcal{C}_{f}$ to be explored, this is performed in the following increments.

First, the added feature combinations ($A' \setminus A$), which do not contain any features from $\mathcal{F}_{(-)}$, are explored.\footnote{
The notation $\mathcal{M} \setminus S$ means that all features in $S$ together with their sub-trees (all children features) are removed from the feature model.}

$$  
\begin{array}{rl}
	\mathcal{C}_{f} \gets & \textrm{getFeatureCombinations}(\mathcal{M}' \setminus \mathcal{F}_{(-)}, f)  \textrm{ } \setminus \\
	  & \textrm{getFeatureCombinations}(\mathcal{M}, f);
\end{array}
  $$
	
	Then, the remaining feature combinations ($A' \cap A$), which do not contain any features from $\mathcal{F}_{(-)}$, are explored.

$$  
\begin{array}{rl}
	\mathcal{C}_{f} \gets & \textrm{getFeatureCombinations}(\mathcal{M}' \setminus \mathcal{F}_{(-)}, f)  \textrm{ } \cap \\
	  & \textrm{getFeatureCombinations}(\mathcal{M}, f);
\end{array}
  $$

And only if all these feature combinations have been explored, the remaining ones are explored.

\section{Experiments}
\label{sec:setup}
This section presents the design and the results of a set of experiments to assess and compare our online learning strategies with the random learning strategy.
The random learning strategy serves as our baseline as it represents existing online learning strategies for rule-based adaptation.

\subsection{Research Questions}

We aim to answer the following research questions:

\textbf{RQ1 (Convergence of feature-model-guided online learning).} \emph{How does the speed of convergence using feature models to explore the adaptation space compare to the speed of convergence using random exploration?}
We aim at determining whether using knowledge about the structure of the adaptation space speeds up convergence when compared with a random learning strategy, thereby helping address the potentially large size of the adaptation space.

\textbf{RQ2 (Convergence of evolution-aware online learning).} \emph{How does the speed of convergence of the evolution-aware learning strategies compare to the convergence of evolution-unaware learning strategies?}
We aim at assessing whether taking into account knowledge about system evolution speeds up convergence of online learning, thereby helping to cope with system evolution.

\textbf{RQ3 (Impact of evolution-aware online learning strategies on system quality).} \emph{What impact on the quality characteristics of a self-adaptive system can be observed when using evolution-aware online learning strategies?}
With this question we aim at understanding how our evolution-aware learning strategies perform when evolving and adapting a system during actual execution.
In particular, we are interested in the effect that evolution-aware learning strategies have on system quality when compared with evolution-unaware learning strategies.
\vspace{-2em}

\subsection{Design}

Our experiments build on four real-world systems and datasets that are listed in \citetable{dataset}.
We purposefully chose the four systems to exhibit different characteristics of the system's adaptation space.
The systems differ with respect to the size of the adaptation space (\ie the number of feature combinations), the number of features and the depth of the feature models.
The feature models of all four systems are provided as supplemental material to this paper.

\begin{table}[h!]
 \centering
     \renewcommand{\arraystretch}{1.5}
\begin{tabular}{| c |  r | r | r |}
\hline
& size of  adap-& number of &  feature model\\
& tation space & features & depth \\
\hline
CloudRM 		&  344  &  63  &  3      \\
BerkeleyJ 		&  360  &  26  &  5        \\
LLVM 		&  1024  &  11  &  1      \\
BerkeleyC         &  2560  &  18  &  2     \\
\hline
\end
{tabular}
\caption{Systems and datasets used for the experiments (ordered by size of adaptation space).}
\label{table:dataset}
\end{table}

\begin{table*}[t]
 \centering
     \renewcommand{\arraystretch}{1.5}
\begin{tabular}{| c | r | r r | r r | r r | r r r}
\hline
   & Baseline (Rand) & \multicolumn{2}{c|}{Inc} & \multicolumn{2}{c|}{Deg} \\
  & nbr explored & nbr explored  & \emph{reduction} & nbr explored  & \emph{reduction} \\

\hline
CloudRM & 174 (50.6\%) & 170 (49.4\%) & \emph{2.3\%} 	& 160 (46.5\%) & \emph{8.0\%  }  \\
BerkeleyJ 				& 186 (51.7\%) & 165 (45.8\%) & \emph{11.3\%} 	& 151 (41.9\%) & \emph{18.8\%}   \\
LLVM 				& 506 (49.4\%) & 508 (49.6\%) & \emph{$-$0.4\%}  	& 503 (49.1\%) & \emph{0.6\%}    \\
BerkeleyC				& 1285 (50.2\%) & 1272 (49.7\%) & \emph{1.0\%}	& 1270 (49.6\%) & \emph{1.2\%}  \\

\hline\end
{tabular}
\caption{Average (and relative) number of feature combinations explored until convergence and reduction compared to baseline.}
\label{table:strategies_results}
\end{table*}

The CloudRM dataset stems from a parametrized cloud resource management system and was created as part of previous work on cloud computing~\cite{MannM17}.
CloudRM controls the allocation of computational tasks to virtual machines and the allocation of virtual machines to physical machines in a cloud data center. 
Moreover, it continuously re-optimizes the placement of virtual machines on physical machines using live migrations to respond to changes in the workload, with the overall aim of minimizing the total energy consumption of the data center while keeping the number of migrations low. 
CloudRM supports multiple algorithms for the selection of a virtual machine for a new task, and the algorithms can be parameterized using different sets of parameters.
Adaptation actions for CloudRM are thus the selection of different algorithms and the parametrization of these algorithms.
We consider energy consumption and number of virtual machine migrations as the specific quality characteristics of interest for the CloudRM system.

The BerkeleyJ, LLVM, and BerkeleyC datasets were collected by Siegmund \etal \cite{Siegmund_ICSE_2012} and were used for experimentation with reconfigurable systems in order to predict their performance.
They describe the reconfigurable open source database systems BerkeleyJ and BerkeleyC, as well as the reconfigurable open source LLVM compiler.
We chose these systems because the datasets include performance measurements for all system configurations, which were measured using standard benchmarks.
As adaptation actions we consider changing at runtime the configurations offered by these systems.
We consider response time as the specific quality characteristic of interest, because this was available across all three datasets.\footnote{We did not consider the other datasets in Siegmund \etal, because in these datasets many of the configurations are associated with the same response time.
This means the chance of finding an effective feature combination is very high, making the learning process converge too fast to observe any differences between the strategies.}

All learning strategies were implemented in Java.
Feature model management and analysis were performed using the FAMA library.\footnote{\url{http://www.isa.us.es/fama/}}
More specifically, we used the FAMA library to identify possible feature combinations from a feature model and reason on partial feature combinations in order to compute the feature degree.

\subsection{Execution}
\label{sec:exec}

To answer \textbf{RQ1}, we use the four datasets as follows.
We \emph{(i)} determine a target quality requirement value by randomly selecting one value from the dataset, and \emph{(ii)} run the learning strategies until they find an effective feature combination, \ie a feature combination that achieves this target value. 

For each strategy, we measure the speed of convergence by counting the number of iterations required to find an effective feature combination.
We also measure the relative speed of convergence by computing the ratio of visited feature combinations over all feature combinations (\ie the size of the adaptation space).
To avoid chance effects, we run the experiment $n$ times, where $n$ is the size of the adaptation space of the respective system, and average the results.

To answer \textbf{RQ2}, we compare the evolution-aware strategies against the evolution-unaware ones using an evolution scenario for each of the systems. 
For each step of the evolution scenario, the experiment measures the speed of convergence.
The measurement procedure follows the one described for RQ1.

For CloudRM, we use the following 4-step evolution scenario.
The scenario starts from an initial system version with a single feature called \mm{Simple} placement, which creates a dedicated virtual machine for each task to be deployed in the cloud. 
Each evolution step adds features for different placement algorithms: 

\begin{enumerate}
\item \mm{Multiple} placement is added, allowing a given number of tasks to be deployed on a virtual machine.
\item \mm{Maxsize} placement is added, creating virtual machines whose size is at most 0.25 times the capacity of the available physical servers. 
When there are multiple virtual machines that can accommodate a new task, a virtual machine is selected using the \mm{First-Fit (FF)} heuristic, selecting the first virtual machine that fits the resource needs of the task. 
\item \mm{Maxsize} placement is made parametrizeable by allowing various maximum virtual machine sizes; two new virtual machine selection heuristics \mm{Best-Fit (BF)} and \mm{Worst-Fit (WF)} are added.
\item \mm{Consolidation\_Friendly} placement is added, selecting a physical machine that can accommodate the given task, and then selecting a virtual machine hosted on the physical machine.
\end{enumerate}

For BerkeleyJ, BerkeleyC, and LLVM, we simulate system evolution by first changing all optional features to mandatory ones, thereby reducing the size of the adaptation space.
Then, we start from this reduced adaptation space and, one by one, randomly change the mandatory features back into the original optional features, thereby incrementally increasing the size of adaptation space. 
This results in an $m$-step evolution scenario, with $m$ being the number of optional features.
We defined evolution scenarios for the BerkeleyJ, BerkeleyC, and LLVM systems with $m = $~7, 7, and 10 respectively.

To answer \textbf{RQ3}, we employ the CloudRM system and measure the quality characteristics ``energy consumption'' and ``number of virtual machine migrations''.
We use the same evolution scenario as for RQ2, but now we actually execute the adaptation actions in the running system.
This means for each of the feature combinations explored by the learning strategies, the system is reconfigured accordingly at runtime.
We measure the impact of these adaptations on system quality for the evolution-aware and evolution-unaware learning strategies.


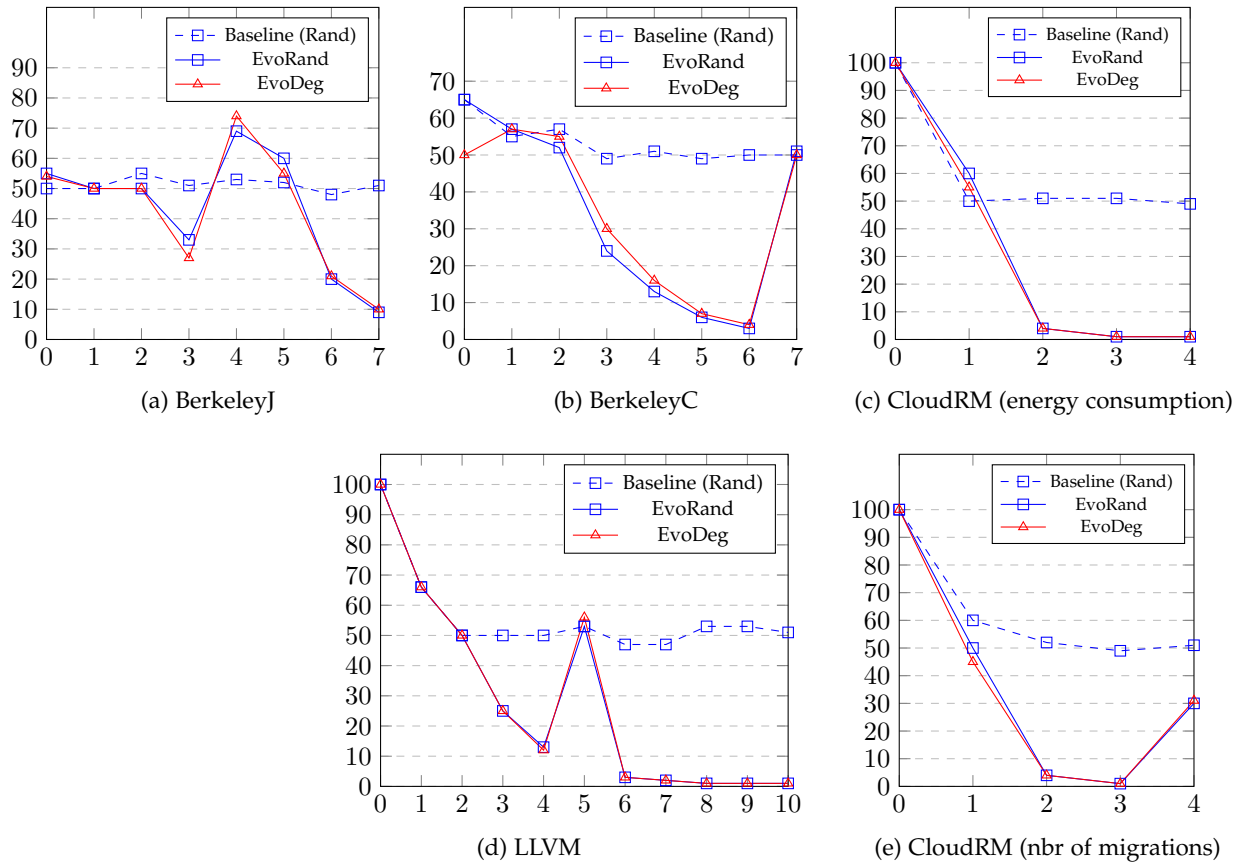
\begin{figure*}[t]  
\centering 
  \begin{subfigure}[b]{0.3\linewidth}
\begin{tikzpicture}
\begin{axis}[
    xmin=0, xmax=7,
    ymin=0, ymax=110,
    xtick={0,1,2,3,4,5,6,7},
    ytick={0,10,20,30,40,50,60,70,80,90},
    legend pos=north east,
    ymajorgrids=true,
    grid style=dashed,
    width=6cm,height=6cm,    
    legend style={nodes={scale=0.75, transform shape}}, 
]    
    \addplot[
    dashed,
    color=blue,
    mark=square,
    mark options=solid,
    ]
    coordinates {
    (0,50)(1,50)(2,55)(3,51)(4,53)(5,52)(6,48)(7,51)
    };
    \addlegendentry{Baseline (Rand)}
    
    \addplot[
    color=blue,
    mark=square,
    mark options=solid,
    ]
    coordinates {
    (0,55)(1,50)(2,50)(3,33)(4,69)(5,60)(6,20)(7,9)
    };
    \addlegendentry{EvoRand}
    
        \addplot[
    color=red,
    mark=triangle,
    ]
    coordinates {
    (0,54)(1,50)(2,50)(3,27)(4,74)(5,55)(6,21)(7,10)
    };
    \addlegendentry{EvoDeg}

\end{axis}
\end{tikzpicture}%
    \caption{BerkeleyJ} \label{fig:Berkeley}  
  \end{subfigure}
\begin{subfigure}[b]{0.3\linewidth}
\begin{tikzpicture}
\begin{axis}[
    xmin=0, xmax=7,
    ymin=0, ymax=90,
    xtick={0,1,2,3,4,5,6,7},
    ytick={0,10,20,30,40,50,60,70},
    legend pos=north east,
    ymajorgrids=true,
    grid style=dashed,
    width=6cm,height=6cm,
    legend style={nodes={scale=0.8, transform shape}}, 
]
 
\addplot[
    dashed,
    color=blue,
    mark=square,
    mark options=solid,
    ]
    coordinates {
	(0,65)(1,55)(2,57)(3,49)(4,51)(5,49)(6,50)(7,50)   
	};
    \addlegendentry{Baseline (Rand)}

\addplot[
    color=blue,
    mark=square,
    ]
    coordinates {
    (0,65)(1,57)(2,52)(3,24)(4,13)(5,6)(6,3)(7,51)   
    };
    \addlegendentry{EvoRand}
    
            \addplot[
    color=red,
    mark=triangle,
    ]
    coordinates {
    (0,50)(1,57)(2,55)(3,30)(4,16)(5,7)(6,4)(7,50)
    };
    \addlegendentry{EvoDeg}

\end{axis}
\end{tikzpicture}
\caption{BerkeleyC} \label{fig:BerkeleyC}  
\end{subfigure}
\begin{subfigure}[b]{0.3\linewidth}
\begin{tikzpicture}
\begin{axis}[
    xmin=0, xmax=4,
    ymin=0, ymax=120,
    xtick={0,1,2,3,4},
    ytick={0,10,20,30,40,50,60,70,80,90,100},
    legend pos=north east,
    ymajorgrids=true,
    grid style=dashed,
    width=5.5cm,height=6cm,
    legend style={nodes={scale=0.7, transform shape}}, 
]
 
\addplot[
    dashed,
    color=blue,
    mark=square,
    mark options=solid,
    ]
    coordinates {
	(0,100)(1,50)(2,51)(3,51)(4,49)
	};
    \addlegendentry{Baseline (Rand)}

\addplot[
    color=blue,
    mark=square,
    ]
    coordinates {
    (0,100)(1,60)(2,4)(3,1)(4,1) 
    };
    \addlegendentry{EvoRand}
    
    \addplot[
    color=red,
    mark=triangle,
    ]
    coordinates {
    (0,100)(1,55)(2,4)(3,1)(4,1) 
    };
    \addlegendentry{EvoDeg}

\end{axis}
\end{tikzpicture}
\caption{CloudRM (energy consumption)} \label{fig:learningCloudRM_energy} 
\end{subfigure}

\vspace{5mm}
\qquad 
\qquad 
\qquad 
\qquad 
\qquad 
\qquad 
\hspace{2mm}
  \begin{subfigure}[b]{0.3\linewidth}
\begin{tikzpicture}
\begin{axis}[
    xmin=0, xmax=10,
    ymin=0, ymax=110,
    xtick={0,1,2,3,4,5,6,7,8,9,10},
    ytick={0,10,20,30,40,50,60,70,80,90,100},
    legend pos=north east,
    ymajorgrids=true,
    grid style=dashed,
    width=7cm,height=6cm,    
    legend style={nodes={scale=0.8, transform shape}}, 
]
 
\addplot[
    dashed,
    color=blue,
    mark=square,
    mark options=solid,
    ]
    coordinates {
	(0,100)(1,66)(2,50)(3,50)(4,50)(5,53)(6,47)(7,47)(8,53)(9,53)(10,51)    
	};
    \addlegendentry{Baseline (Rand)}

\addplot[
    color=blue,
    mark=square,
    ]
    coordinates {
    (0,100)(1,66)(2,50)(3,25)(4,13)(5,53)(6,3)(7,2)(8,1)(9,1)(10,1)    
    };
    \addlegendentry{EvoRand}
    
    \addplot[
    color=red,
    mark=triangle,
    ]
    coordinates {
    (0,100)(1,66)(2,50)(3,25)(4,12)(5,56)(6,3)(7,2)(8,1)(9,1)(10,1)
    };
    \addlegendentry{EvoDeg}

\end{axis}
\end{tikzpicture}
\caption{LLVM} \label{fig:LLVM}  
\end{subfigure}
\qquad 
\qquad
\begin{subfigure}[b]{0.3\linewidth}
\begin{tikzpicture}
\begin{axis}[
    xmin=0, xmax=4,
    ymin=0, ymax=120,
    xtick={0,1,2,3,4},
    ytick={0,10,20,30,40,50,60,70,80,90,100},
    legend pos=north east,
    ymajorgrids=true,
    grid style=dashed,
    width=5.5cm,height=6cm,    
    legend style={nodes={scale=0.7, transform shape}}, 
]    
    \addplot[
    dashed,
    color=blue,
    mark=square,
    mark options=solid,
    ]
    coordinates {
	(0,100)(1,60)(2,52)(3,49)(4,51)
	};
    \addlegendentry{Baseline (Rand)}
    
    \addplot[
    color=blue,
    mark=square,
    ]
    coordinates {
    (0,100)(1,50)(2,4)(3,1)(4,30) 
    };
    \addlegendentry{EvoRand}
    
                \addplot[
    color=red,
    mark=triangle,
    ]
    coordinates {
    (0,100)(1,45)(2,4)(3,1)(4,31) 
    };
    \addlegendentry{EvoDeg}

\end{axis}
\end{tikzpicture}
\caption{CloudRM (nbr of migrations)} 
\label{fig:learningCloudRM_mig}
\end{subfigure}
\caption{Convergence for different steps of the evolution scenarios\\
{(}x-axis: evolution step; y-axis: relative number of feature combinations explored until convergence{)}}
\label{fig:comparison} 
\vspace{-1em}
\end{figure*}

This experiment is based on a real-world workload trace with 10,000 tasks, in total spanning over a time frame of roughly one month \cite{Mann18}. 
To ensure consistency among the results, the same workload was replayed after each step in the evolution scenario. 
CloudRM decides on the placement of new tasks whenever they are entered into the system (as driven by the workload trace).
Additionally, CloudRM re-executes the placement algorithms every five minutes to re-optimize the placement of virtual machines. 
To allow sufficient time in the experiment to observe the impact of the execution of an adaptation action selected by online learning, CloudRM is allowed to run one hour after each adaptation before the next adaptation action is selected.

\subsection{Results}
\label{sec:evaluation}

For \textbf{RQ1}, \citetable{strategies_results} presents the measurements of the convergence speed of the different online learning strategies.
It gives for each system and for each strategy, the average and relative number of feature combinations explored until an effective one is found, as well as the relative reduction of the number of explored feature combinations compared to the baseline random strategy.\footnote{Note that the CloudRM results for energy and migrations are the same, as the strategies explore the feature model in the same order.}

The results suggest that the learning strategies that consider the structure of the feature model (Inc and Deg) perform better than the baseline (Rand) for feature models with greater depth.
Whenever the feature model is \emph{flat}, \ie when the feature model has only few levels, all strategies are very similar in terms of convergence, with around 50\% of the feature combinations explored before finding the target one.\footnote{Results may vary for flat models that contain constraints, since constraints change the feature degree of involved features.}
This is the case for LLVM, which has a depth of 1, and also the case for BerkeleyC, which has a depth of 2.
The reason is that flat models do not provide enough structure for our online learning strategies, and thus they behave like a random learning strategy.

For feature models with greater depth, faster convergence is achieved when considering the structure of the feature model.
This is the case, for instance, for BerkeleyJ with a depth of 5.
Across all systems, the best results are achieved by the strategy with the lowest amount of randomness (Deg).
For Deg, a speed up of convergence of up to 18.8\% for BerkeleyJ was measured. 
This means that the Deg strategy had to explore 18.8\% less adaptation actions than the baseline strategy.

For \textbf{RQ2}, Figure~\ref{fig:comparison} plots the relative speed of convergence for each step of a system's evolution scenario.
In addition, \citetable{strategies_results_rq3} shows the cumulative number of feature combinations explored across all steps of the evolution scenario, as well as their relative reduction.
The two evolution-unaware strategies that performed worst resp. best in RQ1, were Rand resp. Deg.
To measure the improvement of the evolution-aware strategies, we thus use their evolution-aware versions, \ie we use EvoRand and EvoDeg.
Like for RQ1, we use the evolution-unaware random strategy (Rand) as the baseline for comparison. 

\begin{table}[!b]
 \centering
     \renewcommand{\arraystretch}{1.5}
\begin{tabular}{| c | c | r r | r r |}
\hline
& Baseline & \multicolumn{4}{c|}{Evolution-aware}  \\

& Rand & \multicolumn{2}{c|}{EvoRand} &  \multicolumn{2}{c|}{EvoDeg} \\
\hline
\shortstack{CloudRM\\(migrations)}& 261 &   118 & \emph{54.8\%}  	   &  117 & \emph{55.1\%}	  	    \\ 
\shortstack{CloudRM\\(energy)}& 201 &  16  & \emph{92.0\%}      &  15 & \emph{92.5\%}	  	    \\ 
BerkeleyJ & 493 & 274 & \emph{44.4\%}	   & 275 & \emph{44.2\%} 	\\ 
LLVM 				& 454 & 38 & \emph{91.6\%}   &  37 & \emph{91.8\%}	  	\\ 
BerkeleyC         		& 2574 & 1559 & \emph{39.4\%}   & 1535 & \emph{40.3\%}	\\ 
\hline
Average & & & \emph{64.4\%} &&  \emph{64.8\%}\\
\hline
\end
{tabular}
\caption{Cumulative number of feature combinations explored during system evolution and \emph{reduction} (in \%) compared with baseline.}
\label{table:strategies_results_rq3}
\end{table}

As seen in \citetable{strategies_results_rq3}, evolution-aware learning in general shows a strong reduction in the number of feature combinations to be explored before finding an effective one.
These reductions range from 39.4\% (BerkeleyC) to 92.5\% (CloudRM -- energy).
There is only a small difference of 0.4\% on average between the EvoDeg and the EvoRand strategies.
This small difference indicates that first exploring feature combinations added by evolution and exploiting knowledge about whether a feature was part of an effective feature combination in the past has a stronger impact on convergence than how the adaptation space is traversed.

As visible in Figure~\ref{fig:comparison}, in few cases the evolution-aware strategies may take wrong decisions.
In these cases, the speed of convergence may be the same or slightly higher than for the evolution-unaware strategies.
One reason is that even though a feature $f_1$ may have a negative impact on the system's quality in isolation, this feature in combination with a newly introduced feature $f_2$ in the evolved system can have a positive impact.
However, our evolution-aware strategies currently do not consider such feature interactions and thus feature $f_1$ will be only explored after all other feature combinations that do not include this feature have been explored.
Technically, feature $f_1$ is moved to the set $\mathcal{F}_{(-)}$ (see Section~\ref{sec:ExploEvo}).
This is what happened in steps 4 and 5 with BerkeleyJ, step 7 with BerkeleyC and step 5 with LLVM.
Another reason is that $\mathcal{F}_{(+)}$ and $\mathcal{F}_{(-)}$ do not contain many features, because the feature models before the evolution step are very small.
Therefore, there is not much knowledge to be reused across the evolution step.
This is what happened in step 1 for CloudRM (energy).
Here, the feature model at step 0 only includes a single feature, the \mm{Simple} feature.

In Figure~\ref{fig:comparison}, we can also observe the general trend that the more evolution steps a system undergoes, the higher the reduction in the number of feature combinations to be explored becomes.
This suggests that the speed of convergence increases with the number of evolution steps a system undergoes.
This can be explained by the fact that the strategies accumulate knowledge from each of the previous evolutions.
Technically, this means that from each evolution, the strategies learn more precise sets $\mathcal{F}_{(+)}$ and $\mathcal{F}_{(-)}$.

For \textbf{RQ3}, \citetable{savings} shows the percentage of savings in terms of energy consumption and number of virtual machine migrations for each of the four steps of the CloudRM evolution scenario.
We use the worst performing evolution-aware strategy from RQ2 (EvoRand) and compare it with the random baseline (Rand).
As visible in \citetable{savings}, evolution-aware learning leads to considerable savings in terms of the two considered quality characteristics in all but one case.
The negative energy consumption saving for evolution step 1 can be explained by evolution-aware learning exploring more feature combinations than evolution-unaware learning, as explained before in relation to RQ2.

\begin{figure}[t]

\begin{subfigure}[c]{.42\textwidth}
\includegraphics[width=\columnwidth]{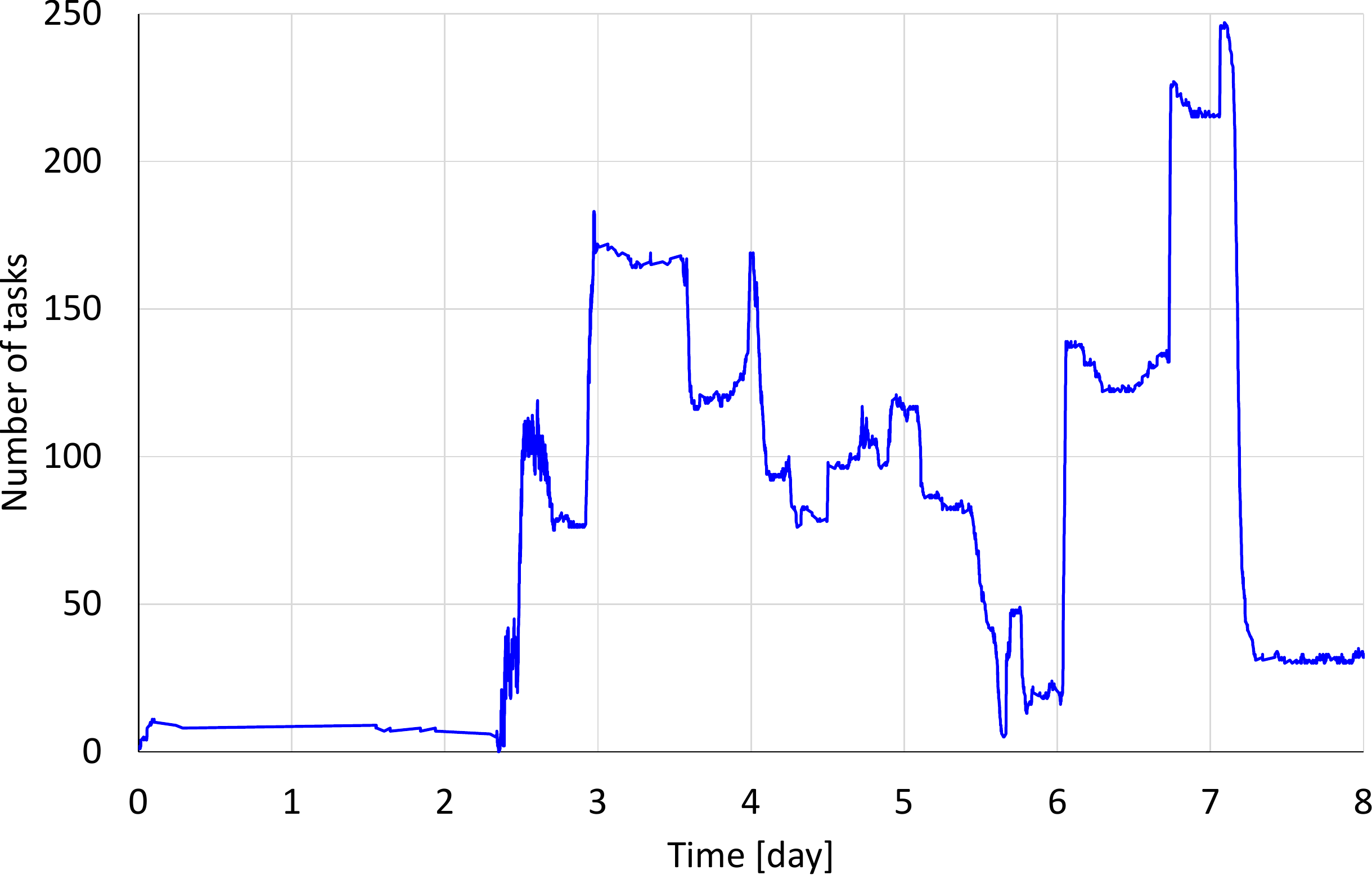}
\subcaption{Workload used for CloudRM experiment}
\vspace{1em}
\end{subfigure}

\begin{subfigure}[c]{.42\textwidth}
\includegraphics[width=\columnwidth]{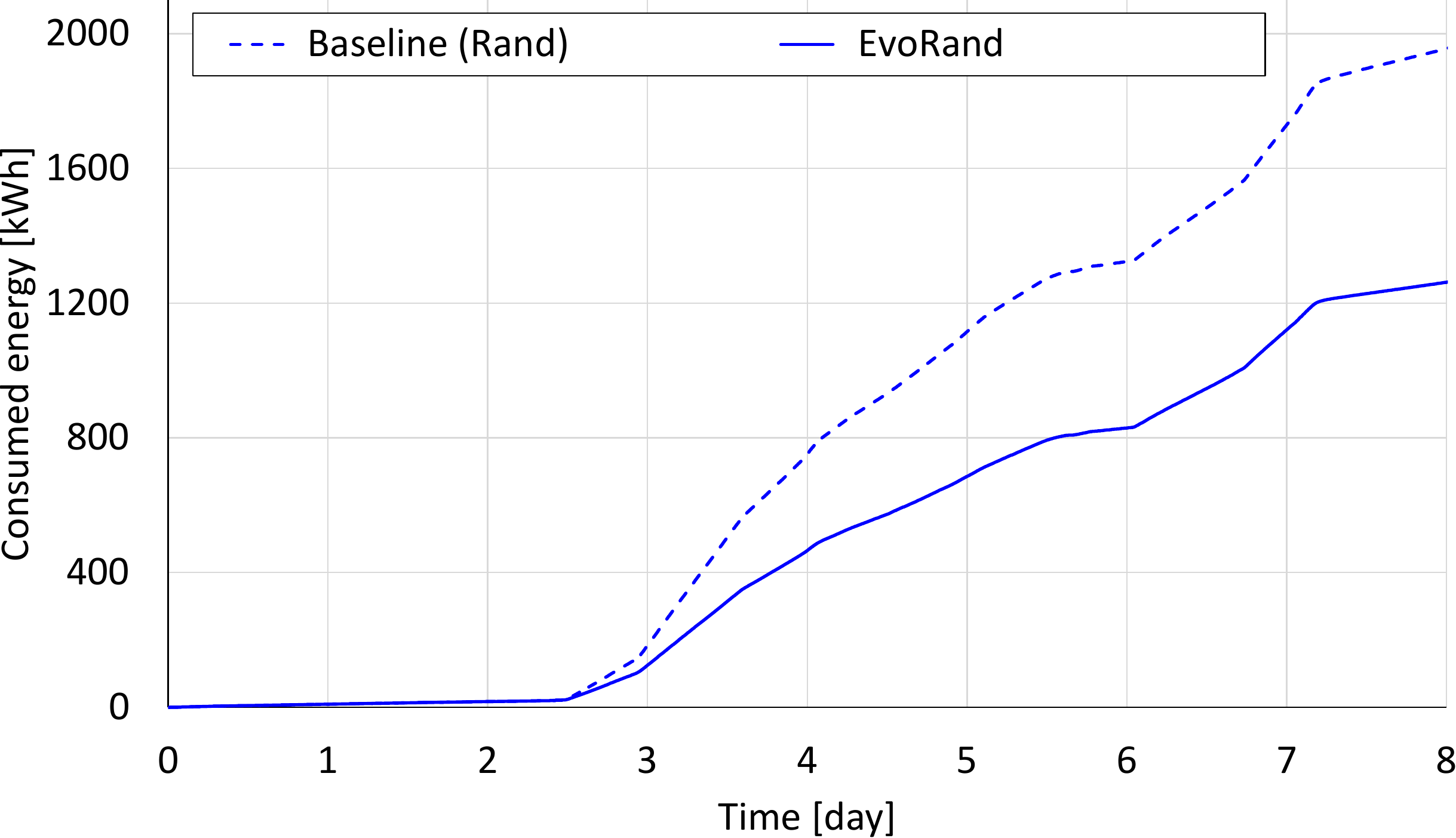}
\subcaption{Energy consumption (evolution step 4)}
\vspace{1em}
\end{subfigure}

\begin{subfigure}[c]{.42\textwidth}
\includegraphics[width=\columnwidth]{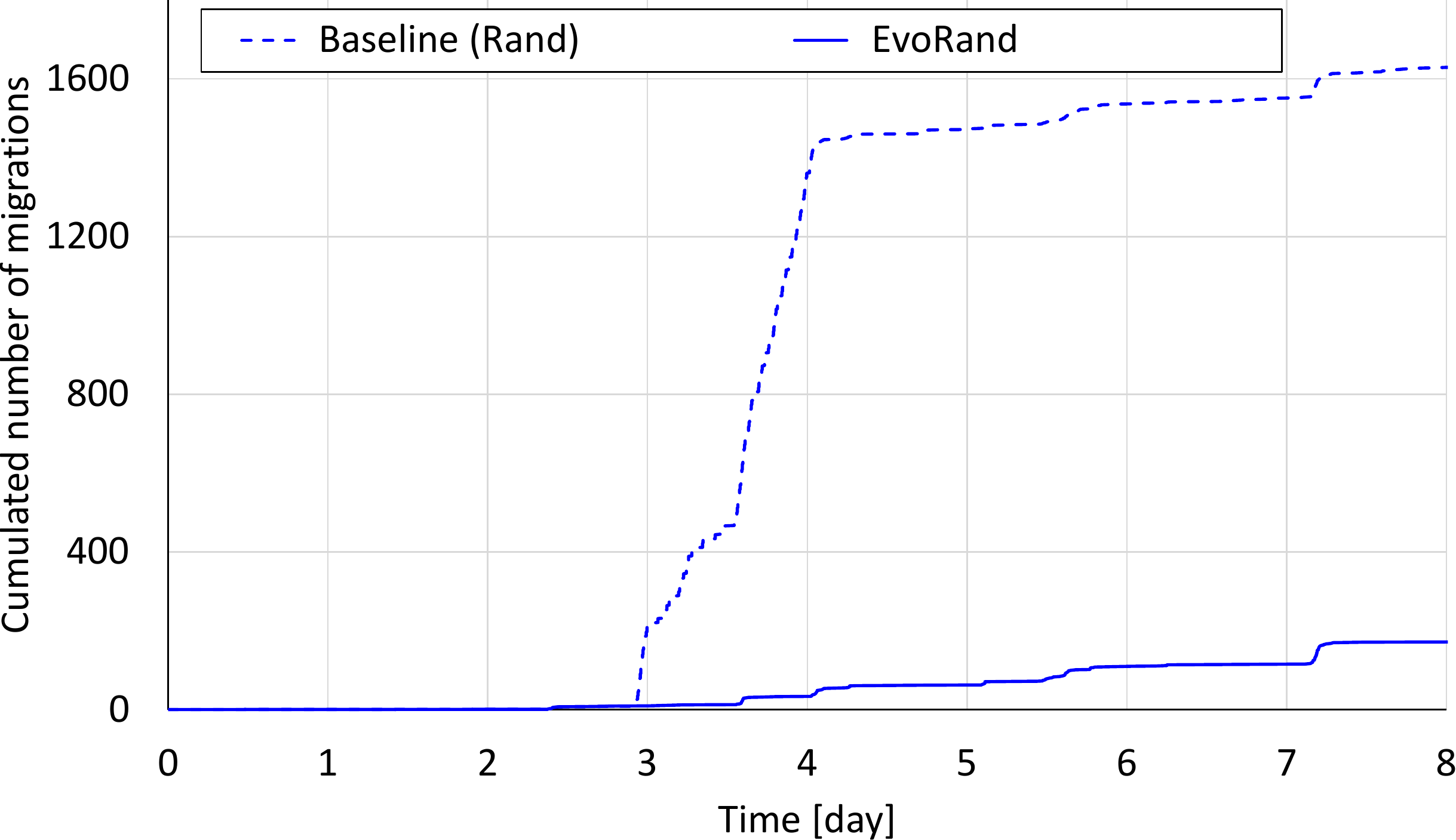}
\subcaption{Migrations (evolution step 4)}
\end{subfigure}

\caption{Savings of evolution-aware learning (CloudRM)}
\label{fig:cloudrm}
\vspace{-1.5em}
\end{figure}

\begin{table}[!b]
\centering
\begin{tabular}{|r|c|c|}
\hline
Evol.\ step & Savings in energy & Savings in migrations \\
\hline
1 & \emph{-7.9\%} & \emph{25.3\%} \\
2 & \emph{11.9\%} & \emph{95.4\%} \\
3 & \emph{~8.3\%} & \emph{86.9\%} \\
4 & \emph{35.5\%} & \emph{89.4\%} \\
\hline
Average & \emph{12.0\%} & \emph{74.3\%} \\
\hline
\end{tabular}
\caption{Savings in energy consumption and number of migrations, achieved through evolution-aware learning for CloudRM.}
\label{table:savings}
\end{table}

When comparing the two quality characteristics, higher savings can be achieved in terms of the number of migrations than in terms of energy consumption. 
This is due to the different placement algorithms in CloudRM having a larger variance in the number of migrations than in energy consumption. 

Even though savings in energy consumption appear not to be very high, it should be noted that, in a large cloud data center, already a modest percentage of energy savings has a considerable impact. 
For example, in a data center with 10,000 physical machines consuming on average 300 Watts, assuming a typical power usage effectiveness of 1.7~\cite{Mann18} and an average electricity price of 0.125 \euro~per kWh~\cite{Mann18}, electricity would cost ca. 15,000 \euro\ per day.
In the case of CloudRM, taking evolution into consideration may save 12\% of energy on average and thus may lead to savings of ca. 1,800 \euro\ per day.

To provide a better understanding of how learning behaves over time, Figure~\ref{fig:cloudrm} plots the cumulative amount of energy consumed and number of migrations performed for evolution step 4.
In our experiments it took 180 iterations for the learning process to converge after an evolution step.
As we used one-hour cycles for adaptation (as explained in Section~\ref{sec:exec}), this means it takes 7.5 days to execute these 180 iterations.
Figure~\ref{fig:cloudrm} therefore reports the results computed based on the first eight days of the workload. 

Figure~\ref{fig:cloudrm} shows that the energy consumption and number of migrations are very low in the first couple of days, followed by increased activity starting from the third day. 
This pattern is a characteristic of the workload (shown at the top of Figure~\ref{fig:cloudrm}), and not related to CloudRM or our learning strategies. 
During the first couple of days, the workload is very low, rising after day 2.
Thus, starting with day 3, evolution-aware learning makes an increasing difference: whilst evolution-unaware learning explores a number of ineffective feature combinations, evolution-aware learning more quickly finds effective feature combinations and thereby converges faster. 
As a result, both energy consumption and the number of migrations grow much slower when using evolution-aware learning. 

\subsection{Threats to Validity}

\textbf{Internal validity.} 
The random baseline strategy as well as our online learning strategies -- even if to a much lesser degree -- exhibit random behavior during adaptation space exploration.
In order to minimize chance effects due to this random behavior, we therefore repeated the experiments $n$ times ($n$ being the size of the adaptation space) and used the average results for comparison of the strategies.

We purposefully focused on evolution steps that increase the size of the adaptation space in order to assess in how far our strategies are able to capture  adaptation spaces of increasingly larger size. 
Our experiments may be complemented by analyzing in how far the strategies differ when the size of the adaptation space is reduced.
Even though in an adaptation space of reduced size, fewer feature combinations have to be explored -- leading to faster convergence  -- there still may be differences in the way these fewer feature combinations are explored.

To measure the speed of convergence, we counted the number iterations  until a feature combination is found that achieves a specific target quality requirement value.
While being an objective metric, this definition of convergence is rather narrow.
Providing a broader definition of convergence, say by giving lower or upper bounds around a target value, may deliver different results.

\textbf{External validity.}
We used four real-world systems from different application domains to measure the speed of convergence of the different strategies.
These four systems also differ in key aspects.
They differ in the shape of their feature models, including the number of features and the depth of the feature model.
Also, they differ in the size of the adaptation space.
Overall, this contributes to the generalizability of our findings with respect to RQ1 and RQ2. 

We used a cloud resource management system to measure the effect of online learning during actual system operation.
Even though we have used a real-world workload trace, results are only for a single system.
This limits generalizability of our findings with respect to RQ3.

\section{Related Work}
\label{sec:related_work}

This section discusses existing online learning techniques for self-adaptive systems and specifically analyzes them with respect to how they address convergence and system evolution.
The discussion is structured along the two main machine learning paradigms used, reinforcement learning and supervised learning, as well as their combination.

\subsection{Reinforcement Learning}
\label{sec:sota_RL}

In reinforcement learning, the system learns the effectiveness of its actions through interactions with its environment~\cite{RL}.
Reinforcement learning can be used to solve sequential decision tasks~\cite{Dietterich96}, where the system aims to maximize its long-term rewards for taking a series of actions in an unknown environment.
The system observes the current environment state and then selects and executes an action, which in turn may cause a change of the environment state. 
The system receives a reward value as feedback for executing an action. 
Reinforcement learning aims to find an action-selection policy that optimizes long-term rewards. 

Amoui \etal propose using Q-Learning and SARSA (two concrete reinforcement learning algorithms) for self-adaptive systems~\cite{amoui2008adaptive}. 
They propose speeding up convergence using offline learning and using simulations of the environment to generate a sufficient number of observations.
They observe that different reinforcement learning algorithms may exhibit different speeds of convergence depending on the concrete application context.
However, they do not take into account additional knowledge about the software system to speed up convergence.
Also, they do not address system evolution.

Kim and Park propose Q-Learning as a concrete algorithm to learn adaptation rules at runtime~\cite{KimP09}.
They propose using goal and scenario models to support a more systematic definition of the reinforcement learning problem (in terms of environment state variables and actions).
They show that online learning may gradually optimize the set of adaptation rules, but provide no further convergence analysis.
They do not address system evolution

Dutreilh \etal employ Q-Learning for autonomic cloud resource management~\cite{dutreilh2011using}.
They experiment with speeding up convergence by providing a good initial estimate for the Q-function (which represents the learned knowledge), as well as by using statistical estimates about the environment behavior.
They indicate that system evolution may imply a change of system performance and sketch an idea on how to detect such drifts in system performance.
Yet, they do not consider that evolution may also introduce or remove adaptation actions.

Barrett \etal propose using Q-Learning for autonomic cloud resource allocation~\cite{BarrettHD13}.
To facilitate convergence, they propose parallel learning.
However, this requires that several systems concerned with the same resource allocation tasks exist and thus can share the information they learn in parallel.
System evolution is not addressed, and in principle could become difficult if the involved systems underwent different forms of evolution in parallel.

Bu \etal employ Q-Learning for the self-configuration of cloud virtual machines and applications~\cite{BuRX13}.
They reduce the action space to a much smaller sub-set using two complementary strategies.
On the one hand, they split the action space into coarse-grained sub-sets and for each of these sub-sets find a representative action using the simplex method.
On the other hand, they encode domain knowledge into the learning process by setting experience-based thresholds for the adaption actions.
Their experimental results indicate that their approach indeed can speed up convergence. 
Still, they do not address system evolution.

Jamshidi \etal and Arabnejad \etal apply fuzzy Q-Learning and fuzzy SARSA to learn fuzzy adaptation rules~\cite{JamshidiEtAl2016,ArabnejadPJE17}.
They observe that the rate of exploration (randomly choosing an action) versus exploitation (using learned knowledge to choose an action) affects convergence.
As an extension of their initial work, they  demonstrate that transfer learning may speed up learning~\cite{Jamshidi_SEAMS_2017}. 
However, transfer learning is beneficial only if observations from the source environment are much cheaper to collect than samples from the target environment.
Their approach does not address system evolution, as it assumes that the set of adaptation actions to be explored is fixed.

Caporuscio \etal propose using two-layer hierarchical reinforcement learning for multi-agent service assembly~\cite{CaporuscioDGM16}.
Two layers of monitoring information serve as input to the learning process: local monitoring information and monitoring information collected by other agents.
They observe that by sharing monitoring information, the learning process converges faster than when learning in isolation.
Like for Barret \etal, this requires that several systems exist that can share monitoring information.
They do not address system evolution.

Filho and Porter~\cite{FilhoP17} use an approach inspired by  reinforcement learning to determine which composition of software components best suits the current environment situation.
Their approach starts with the exhaustive exploration of every possible adaptation action in the adaptation space.
They indicate that this is a clear limitation of their approach for what concerns scalability to large action spaces.
Their learning strategy is unaware of changes in the adaptation space due to system evolution.


Wang \etal combine multi-agent reinforcement learning with game theory for adaptive service compositions~\cite{WangCWYHZB17}.
Their  results indicate that convergence depends on the learning rate (\ie to what degree newly observed rewards override past rewards), the number of agents collaborating, as well as the size of  the adaptation space.
As for Barrett \etal, their approach requires that several systems exists that have the same learning goal.
They do not address how a change in the service composition model due to system evolution may impact on learning.

In our own previous work~\cite{SEAMS16}, we sketch the principal dependencies between online learning and system evolution.
On the one hand, we indicate how feedback from learning  (\eg if no effective feature combination could be found) may trigger system evolution.
On the other hand, we analyze how the adaptation space may change during system evolution and how such a change may affect online learning.
However, we neither provided concrete algorithms nor experimental results for considering system evolution during online learning.
Also, we did not address the issue of convergence in the presence of large adaptations spaces.

\subsection{Supervised Learning}
\label{sec:superv}

In supervised learning, the system learns from a set of labeled training data (\ie input data together with output data).
Supervised learning can be used for one-shot decision tasks~\cite{Dietterich96}, where the input data provides the information available for decision making and the output data describes the correct decisions. 

Esfahani \etal propose an online learning framework that uses feature models to represent the adaptation space~\cite{EsfahaniEM13}.
They learn an analytical model that captures how a feature combination impacts on the system's quality requirements, and use this model for planning  adaptation actions.
To realize their framework, they use the M5 model tree learning algorithm.
They measure the time required to train the model for a given number of observations.
However, they do not measure how many observations are needed to converge to an accurate model.
Also, they do not consider how an evolution of the feature model may impact on the learning process.

Sykes \etal use probabilistic rule learning to update environment models at runtime~\cite{SykesCMKRI13}. 
The environment model, encoded as rules in a logic program, describes what effect adaptation actions have on the environment.
New rules are learned using execution traces  of the running system.
While they evaluate the time required for learning new rules from a set of execution traces, they do not analyze how many execution traces may be required to achieve convergence to a sufficiently accurate environment model.
The impact of system evolution is not considered.

Qian \etal employ case-based reasoning for storing and retrieving adaptation rules~\cite{Qian2015}. 
When facing a new situation, similar cases are retrieved from the case base to find an adaptation rule whose effectiveness has been shown earlier. 
If no similar case can be found in the case base, their approach resorts to using goal models to derive new adaptation rules.
They provide no analysis of the convergence of their approach and whether using goal models to derive new adaptation rules may speed up convergence.
They do not discuss how an update of the goal model due to system evolution would impact the case base.

Quin \etal explicitly address the problem of large adaptation spaces for model-based self-adaptation~\cite{QuinEtAl2019}.
If the adaptation space is large, the resource and time needed for model-based analysis can become prohibitive.
The proposed solution is to employ classification and regression machine learning models to determine a representative and much smaller subset of the adaptation space and only analyze this subset.
To speed up convergence, they use an offline learning phase to train sufficiently accurate machine learning models, which are then updated at runtime.
They do not address system evolution and thus a change of the adaption space.

\subsection{Hybrid Approaches}
\label{sec:hybrid}

Hybrid approaches use reinforcement learning in combination with supervised learning.

Tesauro \etal use Q-Learning in combination with an artificial neural network for autonomic resource allocation in data centers~\cite{TesauroJDB07}.
To capture large spaces of environment states, they use an an artificial neural network in the form of a multi-layer perceptron to approximate the value-function.
In Q-Learning, the value function gives the expected cumulative reward when starting from a given environment state~\cite{RL}.
To facilitate convergence, they perform offline learning using queuing models.
They do not address the problem of large action spaces and whether their chosen function approximation may be applicable.
Also, they do not address system evolution.

Xu \etal employ Q-Learning in combination with artificial neural networks for the automatic configuration of cloud virtual machines and applications~\cite{XuRB12}.
Like Tesauro \etal, they use a multi-layer perceptron to approximate the value-function.
In addition, they perform an an offline learning phase to find a good initial estimate of the action selection policy, thereby facilitating convergence at runtime.
Experimental results indicate that such an offline policy initialization can indeed speed up convergence. 
System evolution is not addressed by their approach.

Moustafa and Zhang propose multi-agent Q-Learning in combination with function approximation via linear regression for adaptive service compositions~\cite{MoustafaZ14}.
To speed up convergence, they propose using collaborative learning, where multiple systems simultaneously explore the set of concrete services to be composed.
They observe that collaborative learning may significantly speed up exploration.
However, like for Barrett \etal (see above), this requires that several systems with the same learning goal exist.
They do not address system evolution.

Zhao \etal propose using reinforcement learning in combination with case-based reasoning to generate and update adaptation rules~\cite{ZhaoZZJ17}.
To populate the case base, they use offline reinforcement learning to learn adaptation rules for different system goals.
At runtime, case-based reasoning is used to select the best fitting rule and reinforcement learning is used to fine-tune this rule.
Their approach may take as long to converge on an optimal rule set as online learning from scratch, but it may start with a higher effectiveness of the adaptation rules.
Even though the approach can handle changes in the priorities of system goals, it does not consider an evolution of the system itself.

\subsection{Summary of Related Work}
\label{sec:sum}

While several approaches in the literature consider how to speed up the convergence of the online learning process, only one approach explicitly addresses the problem of large adaptation spaces~\cite{QuinEtAl2019}.
This approach focuses on reducing the size of the analysis models for model-based adaptation. 
In contrast, we explicitly address the problem of large adaptation spaces for rule-based adaptation by introducing online learning strategies that exploit additional information about the software system in the form of feature models.

System evolution and the impact it may have on the online learning process is not addressed in the literature, except in conceptual form in our own previous work~\cite{SEAMS16}.
We address this gap by introducing evolution-aware online learning strategies for rule-based adaptation.


\section{Conclusion}
\label{sec:conclusion}

We introduced online learning strategies that address potentially large adaptation spaces and that can cope with a change of the adaptation space due to system evolution.
Our online learning strategies use feature models to give structure to the system's adaptation space and thereby guide and speed up the online learning process.

By leveraging the hierarchical structure of feature models, our strategies reduce the amount of randomness when exploring the adaptation space.
The strategies systematically traverse the feature model to select the next adaptation action to be executed and observed.
We thereby address the problem that in the presence of large adaptation spaces, random exploration can lead to slow convergence.

By analyzing the delta between a feature model before and after an evolution step, we make online learning aware of changes in the adaptation space.
Thereby, our strategies can identify added, removed and retained adaptation actions.
We use this information to reuse knowledge across the evolution steps in order not to have to start online learning from scratch, as this would mean knowledge already gained is lost and therefore cannot be used to speed up the convergence of online learning after an evolution step.

Experimental results involving four real-world systems suggest that using feature models to structure the adaptation space can speed up convergence of online learning.
Results indicate that considering the structure of the adaptation space speeds up convergence by up to 18.8\% (with 7.2\% on average).
Additionally considering deltas in the adaptation space due to system evolution speeds up convergence by up to 92.5\% (with 64.6\% on average).
Experimental results for a cloud management system indicate that this faster convergence may lead to energy savings of up to 35.5\% (with 12.0\% on average) and a reduction of virtual machine migrations of up to 89.4\% (with 74.3\% on average).

To conclude, we discuss limitations of our online learning strategies and provide pointers to future work.

\subsection{Limitations}
\label{sec:Limitations}

\emph{Feature interactions.}
We currently do not consider the impact of feature interactions when determining which features to explore in the evolved system.
This means that features that were not part of effective feature combinations may be explored only very late for the evolved system, even though they would lead to an effective feature combination together with the new features. 
Considering such feature interactions may allow improving our evolution-aware online learning strategies.
Existing solutions for feature interaction analysis in software product lines~\cite{ThumAKSS14,MetzgerBLP05} may be used to determine such feature interactions.

\emph{Feature modifications.}
To address system evolution, our strategies analyze the differences in the feature models before and after an evolution step.
Thereby, our strategies can determine feature combinations that were added or removed from the system's adaptation space.
A further possible change introduced during system evolution is the modification of a feature's implementation.
However, such a change is not visible in a feature model.
Encoding such kind of knowledge in the feature models thus could further improve our online learning strategies.

\emph{Adaptation constraints.} 
Our learning strategies are based on the assumption that switching from an active feature combination to any other possible feature combination is always possible, \ie we assume that there are no technical or logical constraints on adaptation. 
On the one hand, this means we are not concerned with the technicalities of how to switch between the feature combinations in the running system.
This is the scope of other work, such as~\cite{ArcegaFHC16,ChenPYNZ14}, which thus may serve to address this concern.
On the other hand, we do not take into account stateful or sequential constraints on adaptation itself.
This means, we can directly switch  between any of the feature combinations in the adaptation space without the need to go through intermediate feature combinations.
However, in reality, only certain paths may be permissible to reach another feature combination from the current one.
To consider adaptation paths, our strategies could be enhanced by building on work such as~\cite{RamirezCMB10,Sousa_SEAMS_2017}.

\emph{Risks of online learning.}
Despite many successful applications of online learning for self-adaptive systems, online learning may not be applicable for all kinds of self-adaptive systems.
Systems may operate in an environment where the trial-and-error nature of online learning may not be tolerable, because adaptation actions may harm their environment~\cite{RL}.
A typical example are safety-critical systems.
For such kinds of systems, online learning may face a too high risk as to be practically applicable.

\subsection{Future Work}
\label{sec:FutureWork}

As part of our future work, we aim to address the current limitations of our online learning strategies.
In particular, this includes relaxing the assumption that switching from one feature combination to any other feature combination is always possible, as well as considering feature interactions and feature modifications during evolution.
In addition, we envision two main extensions:

\emph{Extension to model-based adaptation.} 
We focused on online learning strategies for rule-based adaptation.
Yet, the main ideas underlying our strategies may also be applicable for model-based adaptation.
In model-based adaptation, the aim of online learning is to learn analytical models that facilitate generating effective adaptation actions.
To this end, representative observations of the system and environment need to be collected.
As an important difference to rule-based adaptation, convergence has to be measured differently in model-based adaptation.
Here, the accuracy of the model is a prerequisite for generating effective adaptation actions.
Therefore, it is of interest how fast online learning converges on an accurate model.

\emph{Extension of reinforcement learning.} 
Reinforcement learning is widely used for online learning for self-adaptive systems.
Our online learning strategies may be used to extend existing reinforcement learning algorithms.
As an example, they may be used to augment the exploration phase of reinforcement learning algorithms, such as SARSA or Q-Learning~\cite{RL}.
Instead of randomly selecting the next action during exploration, our strategies may be used to better guide action selection. 
However, existing reinforcement learning algorithms assume that the set of actions remains constant.
Further work is thus required to understand how evolution-aware online learning strategies may be integrated into reinforcement learning algorithms.

  \section*{Acknowledgments}

We cordially thank Amir Molzam Sharifloo for constructive discussions during the conception of this paper, as well as Alexander Palm for his comments on earlier drafts.
Research leading to these results received funding from the European Union's Horizon 2020 research and innovation programme under grant agreements 780351 (ENACT) and 731678 (RestAssured), the European Union's 7th Framework Programme under grant agreement 610802 (CloudWave), and from project EEB - Edifici A Zero Consumo Energetico In Distretti Urbani Intelligenti (Italian Technology Cluster For Smart Communities) - CTN01\_00034\_594053.

\ifCLASSOPTIONcaptionsoff
  \newpage
\fi



\bibliographystyle{IEEEtran}
\bibliography{TSE}

\begin{thebibliography}{10}
\providecommand{\url}[1]{#1}
\csname url@samestyle\endcsname
\providecommand{\newblock}{\relax}
\providecommand{\bibinfo}[2]{#2}
\providecommand{\BIBentrySTDinterwordspacing}{\spaceskip=0pt\relax}
\providecommand{\BIBentryALTinterwordstretchfactor}{4}
\providecommand{\BIBentryALTinterwordspacing}{\spaceskip=\fontdimen2\font plus
\BIBentryALTinterwordstretchfactor\fontdimen3\font minus
  \fontdimen4\font\relax}
\providecommand{\BIBforeignlanguage}[2]{{%
\expandafter\ifx\csname l@#1\endcsname\relax
\typeout{** WARNING: IEEEtran.bst: No hyphenation pattern has been}%
\typeout{** loaded for the language `#1'. Using the pattern for}%
\typeout{** the default language instead.}%
\else
\language=\csname l@#1\endcsname
\fi
#2}}
\providecommand{\BIBdecl}{\relax}
\BIBdecl

\bibitem{SESAS_II}
R.~de~Lemos \emph{et~al.}, ``\BIBforeignlanguage{English}{{Software Engineering
  for Self-Adaptive Systems: A Second Research Roadmap}},'' in
  \emph{\BIBforeignlanguage{English}{Softw. Eng. for Self-Adaptive Systems
  II}}, ser. LNCS.\hskip 1em plus 0.5em minus 0.4em\relax Springer, 2013, vol.
  7475, pp. 1--32.

\bibitem{SalehieT09}
M.~Salehie and L.~Tahvildari, ``Self-adaptive software: Landscape and research
  challenges,'' \emph{{TAAS}}, vol.~4, no.~2, 2009.

\bibitem{BaresiNG06}
L.~Baresi, E.~D. Nitto, and C.~Ghezzi, ``Toward open-world software: Issue and
  challenges,'' \emph{{IEEE} Computer}, vol.~39, no.~10, pp. 36--43, 2006.

\bibitem{KleinMAH14}
C.~Klein, M.~Maggio, K.~{\AA}rz{\'{e}}n, and F.~Hern{\'{a}}ndez{-}Rodriguez,
  ``Brownout: building more robust cloud applications,'' in \emph{36th Intl
  Conf. on Softw. Eng., {ICSE} '14, Hyderabad, India - May 31 - June 07,
  2014}.\hskip 1em plus 0.5em minus 0.4em\relax {ACM}, 2014, pp. 700--711.

\bibitem{KephartC03}
J.~O. Kephart and D.~M. Chess, ``The vision of autonomic computing,''
  \emph{{IEEE} Computer}, vol.~36, no.~1, pp. 41--50, 2003.

\bibitem{IglesiaW15}
D.~G. de~la Iglesia and D.~Weyns, ``{MAPE-K} formal templates to rigorously
  design behaviors for self-adaptive systems,'' \emph{{TAAS}}, vol.~10, no.~3,
  pp. 15:1--15:31, 2015.

\bibitem{ChenB17}
T.~Chen and R.~Bahsoon, ``Self-adaptive and online qos modeling for cloud-based
  software services,'' \emph{{IEEE} Trans. Software Eng.}, vol.~43, no.~5, pp.
  453--475, 2017.

\bibitem{JamshidiPM16}
P.~Jamshidi, C.~Pahl, and N.~C. Mendon{\c{c}}a, ``Managing uncertainty in
  autonomic cloud elasticity controllers,'' \emph{{IEEE} Cloud Computing},
  vol.~3, no.~3, pp. 50--60, 2016.

\bibitem{DIppolitoBKMSU14}
N.~D'Ippolito, V.~A. Braberman, J.~Kramer, J.~Magee, D.~Sykes, and S.~Uchitel,
  ``Hope for the best, prepare for the worst: multi-tier control for adaptive
  systems,'' in \emph{36th Intl Conf. on Softw. Eng., {ICSE} '14, Hyderabad,
  India - May 31 - June 07, 2014}.\hskip 1em plus 0.5em minus 0.4em\relax
  {ACM}, 2014, pp. 688--699.

\bibitem{EsfahaniEM13}
N.~Esfahani, A.~Elkhodary, and S.~Malek, ``{A Learning-Based Framework for
  Engineering Feature-Oriented Self-Adaptive Software Systems},'' \emph{IEEE
  Trans. Softw. Eng.}, vol.~39, no.~11, pp. 1467--1493, Nov 2013.

\bibitem{Esfahani_ESEC_2011}
N.~Esfahani, E.~Kouroshfar, and S.~Malek, ``Taming uncertainty in self-adaptive
  software,'' in \emph{19th Symposium on the Foundations of Softw. Eng. / 13th
  European Softw. Eng. Conf., {ESEC/FSE} 2013, Szeged, Hungary, September 5-9,
  2011}.\hskip 1em plus 0.5em minus 0.4em\relax {ACM}, 2011, pp. 234--244.

\bibitem{MorenoCGS18}
G.~A. Moreno, J.~C{\'{a}}mara, D.~Garlan, and B.~R. Schmerl, ``Flexible and
  efficient decision-making for proactive latency-aware self-adaptation,''
  \emph{{TAAS}}, vol.~13, no.~1, pp. 3:1--3:36, 2018.

\bibitem{TajalliGEM10}
H.~Tajalli, J.~Garcia, G.~Edwards, and N.~Medvidovic, ``{PLASMA:} a plan-based
  layered architecture for software model-driven adaptation,'' in \emph{25th
  Intl Conf. on Automated Softw. Eng., {ASE2010}, Antwerp, Belgium, September
  20-24, 2010}.\hskip 1em plus 0.5em minus 0.4em\relax {ACM}, 2010, pp.
  467--476.

\bibitem{FlochHSELG06}
J.~Floch, S.~O. Hallsteinsen, E.~Stav, F.~Eliassen, K.~Lund, and E.~Gj{\o}rven,
  ``Using architecture models for runtime adaptability,'' \emph{{IEEE}
  Software}, vol.~23, no.~2, pp. 62--70, 2006.

\bibitem{JamshidiEtAl2019}
P.~Jamshidi, J.~Camara, B.~Schmerl, C.~K{\"a}stner, and D.~Garlan, ``Machine
  learning meets quantitative planning: Enabling self-adaptation in autonomous
  robots,'' in \emph{14th Intl Symposium on Softw. Eng. for Adaptive and
  Self-Managing Systems, {SEAMS} 2019, Montreal, Canada, May 25-26,
  2019}.\hskip 1em plus 0.5em minus 0.4em\relax {ACM}, 2019.

\bibitem{FrommgenRLB15}
A.~Fr{\"{o}}mmgen, R.~Rehner, M.~Lehn, and A.~P. Buchmann, ``Fossa: Learning
  {ECA} rules for adaptive distributed systems,'' in \emph{Intl Conf. on
  Autonomic Computing, {ICAC 2015}, Grenoble, France, July 7-10, 2015}.\hskip
  1em plus 0.5em minus 0.4em\relax {IEEE} Comp. Soc., 2015, pp. 207--210.

\bibitem{LaneseBM10}
I.~Lanese, A.~Bucchiarone, and F.~Montesi, ``A framework for rule-based dynamic
  adaptation,'' in \emph{Trustworthly Global Computing - 5th International
  Symposium, {TGC} 2010, Munich, Germany, February 24-26, 2010, Revised
  Selected Papers}, ser. LNCS, vol. 6084.\hskip 1em plus 0.5em minus
  0.4em\relax Springer, 2010, pp. 284--300.

\bibitem{GarlanCHSS04}
D.~Garlan, S.~Cheng, A.~Huang, B.~R. Schmerl, and P.~Steenkiste, ``Rainbow:
  Architecture-based self-adaptation with reusable infrastructure,''
  \emph{{IEEE} Computer}, vol.~37, no.~10, pp. 46--54, 2004.

\bibitem{RamirezJC12}
A.~J. Ramirez, A.~C. Jensen, and B.~H.~C. Cheng, ``A taxonomy of uncertainty
  for dynamically adaptive systems,'' in \emph{7th Intl Symposium on Softw.
  Eng. for Adaptive and Self-Managing Systems, {SEAMS 2012}, Zurich,
  Switzerland, June 4-5, 2012}.\hskip 1em plus 0.5em minus 0.4em\relax {IEEE}
  Comp. Soc., 2012, pp. 99--108.

\bibitem{FredericksDC14}
E.~M. Fredericks, B.~DeVries, and B.~H.~C. Cheng, ``Autorelax: automatically
  relaxing a goal model to address uncertainty,'' \emph{Empirical Software
  Engineering}, vol.~19, no.~5, pp. 1466--1501, 2014.

\bibitem{SykesCMKRI13}
D.~Sykes, D.~Corapi, J.~Magee, J.~Kramer, A.~Russo, and K.~Inoue, ``Learning
  revised models for planning in adaptive systems,'' in \emph{35th Intl Conf.
  on Softw. Eng., {ICSE} '13, San Francisco, CA, USA, May 18-26, 2013}.\hskip
  1em plus 0.5em minus 0.4em\relax {IEEE} Comp. Soc., 2013, pp. 63--71.

\bibitem{ZhaoZZJ17}
T.~Zhao, W.~Zhang, H.~Zhao, and Z.~Jin, ``A reinforcement learning-based
  framework for the generation and evolution of adaptation rules,'' in
  \emph{Intl Conf. on Autonomic Computing, {ICAC 2017} 2017, Columbus, OH, USA,
  July 17-21, 2017}.\hskip 1em plus 0.5em minus 0.4em\relax {IEEE} Comp. Soc.,
  2017, pp. 103--112.

\bibitem{FilhoP17}
R.~V.~R. Filho and B.~Porter, ``Defining emergent software using continuous
  self-assembly, perception, and learning,'' \emph{{TAAS}}, vol.~12, no.~3, pp.
  16:1--16:25, 2017.

\bibitem{Qian2015}
W.~Qian, X.~Peng, B.~Chen, J.~Mylopoulos, H.~Wang, and W.~Zhao, ``Rationalism
  with a dose of empiricism: combining goal reasoning and case-based reasoning
  for self-adaptive software systems,'' \emph{Requirements Engineering},
  vol.~20, no.~3, pp. 233--252, 2015.

\bibitem{Domingos12}
P.~M. Domingos, ``A few useful things to know about machine learning,''
  \emph{Commun. {ACM}}, vol.~55, no.~10, pp. 78--87, 2012.

\bibitem{LeCunBH15}
Y.~LeCun, Y.~Bengio, and G.~E. Hinton, ``Deep learning,'' \emph{Nature}, vol.
  521, no. 7553, pp. 436--444, 2015.

\bibitem{Tesauro07}
G.~Tesauro, ``Reinforcement learning in autonomic computing: {A} manifesto and
  case studies,'' \emph{{IEEE} Internet Computing}, vol.~11, no.~1, pp. 22--30,
  2007.

\bibitem{MoustafaZ14}
A.~Moustafa and M.~Zhang, ``Learning efficient compositions for qos-aware
  service provisioning,'' in \emph{2014 {IEEE} International Conference on Web
  Services, ICWS, 2014, Anchorage, AK, USA, June 27 - July 2, 2014}.\hskip 1em
  plus 0.5em minus 0.4em\relax {IEEE} Comp. Soc., 2014, pp. 185--192.

\bibitem{KaelblingLM96}
L.~P. Kaelbling, M.~L. Littman, and A.~W. Moore, ``Reinforcement learning: {A}
  survey,'' \emph{J. Artif. Intell. Res.}, vol.~4, pp. 237--285, 1996.

\bibitem{amoui2008adaptive}
M.~Amoui, M.~Salehie, S.~Mirarab, and L.~Tahvildari, ``Adaptive action
  selection in autonomic software using reinforcement learning,'' in \emph{4th
  Intl Conf. on Autonomic and Autonomous Systems, {ICAS}'08, 16-21 March 2008,
  Gosier, Guadeloupe}.\hskip 1em plus 0.5em minus 0.4em\relax IEEE, 2008, pp.
  175--181.

\bibitem{JamshidiEtAl2016}
P.~Jamshidi, A.~Sharifloo, C.~Pahl, H.~Arabnejad, A.~Metzger, and G.~Estrada,
  ``Fuzzy self-learning controllers for elasticity management in dynamic cloud
  architectures,'' in \emph{12th Intl Conf. on Quality of Softw. Architectures,
  QoSA 2016, Venice, Italy, April 5-8, 2016}.\hskip 1em plus 0.5em minus
  0.4em\relax IEEE, 2016, pp. 70--79.

\bibitem{Jamshidi_SEAMS_2017}
P.~Jamshidi, M.~Velez, C.~K\"{a}stner, N.~Siegmund, and P.~Kawthekar,
  ``Transfer learning for improving model predictions in highly configurable
  software,'' in \emph{12th Intl Symposium on Softw. Eng. for Adaptive and
  Self-Managing Systems, {SEAMS 2017}, Buenos Aires, Argentina, May 22-23,
  2017}.\hskip 1em plus 0.5em minus 0.4em\relax IEEE Press, 2017, pp. 31--41.

\bibitem{TesauroJDB07}
G.~Tesauro, N.~K. Jong, R.~Das, and M.~N. Bennani, ``On the use of hybrid
  reinforcement learning for autonomic resource allocation,'' \emph{Cluster
  Computing}, vol.~10, no.~3, pp. 287--299, 2007.

\bibitem{MirandolaPS14}
R.~Mirandola, P.~Potena, and P.~Scandurra, ``Adaptation space exploration for
  service-oriented applications,'' \emph{Sci. Comput. Program.}, vol.~80, pp.
  356--384, 2014.

\bibitem{AlferezPMSD14}
G.~H. Alf{\'{e}}rez, V.~Pelechano, R.~Mazo, C.~Salinesi, and D.~Diaz, ``Dynamic
  adaptation of service compositions with variability models,'' \emph{Journal
  of Systems and Software}, vol.~91, pp. 24--47, 2014.

\bibitem{WeynsI0M18}
D.~Weyns, M.~U. Iftikhar, D.~Hughes, and N.~Matthys, ``Applying
  architecture-based adaptation to automate the management of
  internet-of-things,'' in \emph{12th European Conf. on Softw. Architecture,
  {ECSA} 2018, Madrid, Spain, September 24-28, 2018, Proceedings}, ser. LNCS,
  vol. 11048.\hskip 1em plus 0.5em minus 0.4em\relax Springer, 2018, pp.
  49--67.

\bibitem{KimP09}
D.~Kim and S.~Park, ``Reinforcement learning-based dynamic adaptation planning
  method for architecture-based self-managed software,'' in \emph{{ICSE}
  Workshop on Software Engineering for Adaptive and Self-Managing Systems,
  {SEAMS} 2009, Vancouver, BC, Canada, May 18-19, 2009}, 2009, pp. 76--85.

\bibitem{dutreilh2011using}
X.~Dutreilh, S.~Kirgizov, O.~Melekhova, J.~Malenfant, N.~Rivierre, and
  I.~Truck, ``Using reinforcement learning for autonomic resource allocation in
  clouds: towards a fully automated workflow,'' in \emph{7th Intl Conf. on
  Autonomic and Autonomous Systems (ICAS 2011), Venice/Mestre, Italy, May
  22-27, 2011}, 2011, pp. 67--74.

\bibitem{BarrettHD13}
E.~Barrett, E.~Howley, and J.~Duggan, ``Applying reinforcement learning towards
  automating resource allocation and application scalability in the cloud,''
  \emph{Concurrency and Computation: Practice and Experience}, vol.~25, no.~12,
  pp. 1656--1674, 2013.

\bibitem{BuRX13}
X.~Bu, J.~Rao, and C.~Xu, ``Coordinated self-configuration of virtual machines
  and appliances using a model-free learning approach,'' \emph{{IEEE} Trans.
  Parallel Distrib. Syst.}, vol.~24, no.~4, pp. 681--690, 2013.

\bibitem{CaporuscioDGM16}
M.~Caporuscio, M.~D'Angelo, V.~Grassi, and R.~Mirandola, ``Reinforcement
  learning techniques for decentralized self-adaptive service assembly,'' in
  \emph{5th European Conference on Service-Oriented and Cloud Computing,
  {ESOCC} 2016, Vienna, Austria, September 5-7, 2016, Proceedings}, ser. LNCS,
  vol. 9846.\hskip 1em plus 0.5em minus 0.4em\relax Springer, 2016, pp. 53--68.

\bibitem{WangCWYHZB17}
H.~Wang, X.~Chen, Q.~Wu, Q.~Yu, X.~Hu, Z.~Zheng, and A.~Bouguettaya,
  ``Integrating reinforcement learning with multi-agent techniques for adaptive
  service composition,'' \emph{{TAAS}}, vol.~12, no.~2, pp. 8:1--8:42, 2017.

\bibitem{XuRB12}
C.~Xu, J.~Rao, and X.~Bu, ``{URL:} {A} unified reinforcement learning approach
  for autonomic cloud management,'' \emph{J. Parallel Distrib. Comput.},
  vol.~72, no.~2, pp. 95--105, 2012.

\bibitem{QuinEtAl2019}
F.~Quin, D.~Weyns, T.~Bamelis, S.~{Singh Buttar}, and S.~Michiels, ``Efficient
  analysis of large adaptation spaces in self-adaptive systems using machine
  learning,'' in \emph{14th Intl Symposium on Softw. Eng. for Adaptive and
  Self-Managing Systems, {SEAMS} 2019, Montreal, Canada, May 25-26,
  2019}.\hskip 1em plus 0.5em minus 0.4em\relax {ACM}, 2019.

\bibitem{RL}
R.~S. Sutton and A.~G. Barto, \emph{Reinforcement Learning: An Introduction},
  2nd~ed.\hskip 1em plus 0.5em minus 0.4em\relax Cambridge, MA, USA: MIT Press,
  2018.

\bibitem{DSPL-Ghezzi}
C.~Ghezzi and A.~M. Sharifloo, ``Dealing with non-functional requirements for
  adaptive systems via dynamic software product-lines,'' in \emph{Software
  Engineering for Self-Adaptive Systems {II}}, ser. LNCS, vol. 7475.\hskip 1em
  plus 0.5em minus 0.4em\relax Springer, 2010, pp. 191--213.

\bibitem{KinneerCWGG18}
C.~Kinneer, Z.~Coker, J.~Wang, D.~Garlan, and C.~{Le Goues}, ``Managing
  uncertainty in self-adaptive systems with plan reuse and stochastic search,''
  in \emph{13th Intl Symposium on Softw. Eng. for Adaptive and Self-Managing
  Systems, {SEAMS 2018}, Gothenburg, Sweden, May 28-29, 2018}.\hskip 1em plus
  0.5em minus 0.4em\relax {ACM}, 2018, pp. 40--50.

\bibitem{SEAMS16}
A.~M. Sharifloo, A.~Metzger, C.~Quinton, L.~Baresi, and K.~Pohl, ``Learning and
  evolution in dynamic software product lines,'' in \emph{11th Intl Symposium
  on Softw. Eng. for Adaptive and Self-Managing Systems, {SEAMS} 2016, Austin,
  Texas, USA, May 14-22, 2016}.\hskip 1em plus 0.5em minus 0.4em\relax {ACM},
  2016, pp. 158--164.

\bibitem{EvolvingDSPL-SPLC15}
C.~Quinton, R.~Rabiser, M.~Vierhauser, P.~Gr{\"{u}}nbacher, and L.~Baresi,
  ``Evolution in dynamic software product lines: challenges and perspectives,''
  in \emph{19th Intl Conference on Softw. Product Lines, {SPLC 2015},
  Nashville, TN, USA, July 20-24, 2015}, 2015, pp. 126--130.

\bibitem{Gilles12}
G.~Perrouin, B.~Morin, F.~Chauvel, F.~Fleurey, J.~Klein, Y.~Le~Traon,
  O.~Barais, and J.-M. J{\'e}z{\'e}quel, ``Towards flexible evolution of
  dynamically adaptive systems,'' in \emph{34th Intl Conf. on Softw. Eng.,
  {ICSE 2012}, Zurich, Switzerland, June 4-5, 2012}.\hskip 1em plus 0.5em minus
  0.4em\relax Piscataway, NJ, USA: IEEE Press, 2012, pp. 1353--1356.

\bibitem{Ghezzi17}
C.~Ghezzi, ``Of software and change,'' \emph{Journal of Software: Evolution and
  Process}, vol.~29, no.~9, 2017.

\bibitem{IGI_Agile_Book}
A.~Metzger and E.~{Di Nitto}, ``Addressing highly dynamic changes in
  service-oriented systems: Towards agile evolution and adaptation,'' in
  \emph{Agile and Lean Service-Oriented Development: Foundations, Theory and
  Practice}.\hskip 1em plus 0.5em minus 0.4em\relax IGI Global, 2012, pp.
  33--46.

\bibitem{Bosch-SEAMS16}
J.~Bosch and H.~H. Olsson, ``Data-driven continuous evolution of smart
  systems,'' in \emph{11th Intl Symposium on Softw. Eng. for Adaptive and
  Self-Managing Systems, {SEAMS} 2016, Austin, Texas, USA, May 14-22,
  2016}.\hskip 1em plus 0.5em minus 0.4em\relax {ACM}, 2016, pp. 28--34.

\bibitem{SugiyamaK12}
M.~Sugiyama and M.~Kawanabe, \emph{Machine Learning in Non-Stationary
  Environments - Introduction to Covariate Shift Adaptation}, ser. Adaptive
  computation and machine learning.\hskip 1em plus 0.5em minus 0.4em\relax
  {MIT} Press, 2012.

\bibitem{variability2013}
R.~Capilla, J.~Bosch, and K.~C. Kang, Eds., \emph{Systems and Software
  Variability Management, Concepts, Tools and Experiences}.\hskip 1em plus
  0.5em minus 0.4em\relax Springer, 2013.

\bibitem{SPL-FOSE14}
A.~Metzger and K.~Pohl, ``Software product line engineering and variability
  management: Achievements and challenges,'' in \emph{Future of Software
  Engineering, {FOSE} 2014, Hyderabad, India, May 31 - June 7, 2014}, 2014, pp.
  70--84.

\bibitem{HincheyPS12}
M.~Hinchey, S.~Park, and K.~Schmid, ``Building dynamic software product
  lines,'' \emph{{IEEE} Computer}, vol.~45, no.~10, pp. 22--26, 2012.

\bibitem{MetzgerHPSS07}
A.~Metzger, P.~Heymans, K.~Pohl, P.-Y. Schobbens, and G.~Saval,
  ``Disambiguating the documentation of variability in software product lines:
  A separation of concerns, formalization and automated analysis,'' in
  \emph{15th Intl Requirements Eng. Conf., {RE 2007}, October 15-19th, 2007,
  New Delhi, India}.\hskip 1em plus 0.5em minus 0.4em\relax IEEE, 2007, pp.
  243--253.

\bibitem{MorinBJFS09}
B.~Morin, O.~Barais, J.~J{\'{e}}z{\'{e}}quel, F.~Fleurey, and A.~Solberg,
  ``Models{@}run.time to support dynamic adaptation,'' \emph{{IEEE} Computer},
  vol.~42, no.~10, pp. 44--51, 2009.

\bibitem{GomaaH07}
H.~Gomaa and M.~Hussein, ``Model-based software design and adaptation,'' in
  \emph{2007 {ICSE} Workshop on Softw. Eng. for Adaptive and Self-Managing
  Systems, {SEAMS} 2007, Minneapolis Minnesota, USA, May 20-26, 2007}, 2007,
  pp. 7--16.

\bibitem{CalinescuGKM12}
R.~Calinescu, C.~Ghezzi, M.~Z. Kwiatkowska, and R.~Mirandola, ``Self-adaptive
  software needs quantitative verification at runtime,'' \emph{Commun. {ACM}},
  vol.~55, no.~9, pp. 69--77, 2012.

\bibitem{BenavidesTC05}
D.~Benavides, P.~T. Mart{\'{\i}}n{-}Arroyo, and A.~R. Cort{\'{e}}s, ``Automated
  reasoning on feature models,'' in \emph{17th Intl Conf. on Advanced
  Information Systems Engineering, {CAiSE} 2005, Porto, Portugal, June 13-17,
  2005, Proceedings}, ser. LNCS, vol. 3520.\hskip 1em plus 0.5em minus
  0.4em\relax Springer, 2005, pp. 491--503.

\bibitem{Benavides_2010}
D.~Benavides, S.~Segura, and A.~Ruiz-Cort{\'e}s, ``Automated analysis of
  feature models 20 years later: A literature review,'' \emph{Inf. Syst.},
  vol.~35, no.~6, pp. 615--636, Sep. 2010.

\bibitem{Thum_SPLC_11}
T.~Thum, C.~Kastner, S.~Erdweg, and N.~Siegmund, ``Abstract features in feature
  modeling,'' in \emph{15th Intl Conf. on Software Product Lines, {SPLC} 2011,
  Munich, Germany, August 22-26, 2011}.\hskip 1em plus 0.5em minus 0.4em\relax
  IEEE Comp. Soc., 2011, pp. 191--200.

\bibitem{MannM17}
Z.~{\'{A}}. Mann and A.~Metzger, ``Optimized cloud deployment of multi-tenant
  software considering data protection concerns,'' in \emph{17th Intl Symposium
  on Cluster, Cloud and Grid Computing, {CCGrid} 2017, Madrid, Spain,
  2017}.\hskip 1em plus 0.5em minus 0.4em\relax {IEEE} Comp. Soc., 2017, pp.
  609--618.

\bibitem{Siegmund_ICSE_2012}
N.~Siegmund, S.~S. Kolesnikov, C.~K\"{a}stner, S.~Apel, D.~Batory,
  M.~Rosenm\"{u}ller, and G.~Saake, ``{Predicting Performance via Automated
  Feature-interaction Detection},'' in \emph{34th Intl Conf. on Softw. Eng.,
  ICSE 2012, Zurich, Switzerland, June 4-5, 2012}.\hskip 1em plus 0.5em minus
  0.4em\relax IEEE Press, 2012, pp. 167--177.

\bibitem{Mann18}
Z.~{\'{A}}. Mann, ``Resource optimization across the cloud stack,'' \emph{IEEE
  Transactions on Parallel and Distributed Systems}, vol.~29, no.~1, pp.
  169--182, 2018.

\bibitem{Dietterich96}
T.~G. Dietterich, ``Machine learning,'' \emph{{ACM} Comput. Surv.}, vol.~28,
  no. 4es, p.~3, 1996.

\bibitem{ArabnejadPJE17}
H.~Arabnejad, C.~Pahl, P.~Jamshidi, and G.~Estrada, ``A comparison of
  reinforcement learning techniques for fuzzy cloud auto-scaling,'' in
  \emph{17th Intl Symposium on Cluster, Cloud and Grid Computing, {CCGRID}
  2017, Madrid, Spain, May 14-17, 2017}.\hskip 1em plus 0.5em minus 0.4em\relax
  {IEEE} Comp. Soc., 2017, pp. 64--73.

\bibitem{ThumAKSS14}
T.~Th{\"{u}}m, S.~Apel, C.~K{\"{a}}stner, I.~Schaefer, and G.~Saake, ``A
  classification and survey of analysis strategies for software product
  lines,'' \emph{{ACM} Comput. Surv.}, vol.~47, no.~1, pp. 6:1--6:45, 2014.

\bibitem{MetzgerBLP05}
A.~Metzger, S.~B{\"u}hne, K.~Lauenroth, and K.~Pohl, ``Considering feature
  interactions in product lines: Towards the automatic derivation of
  dependencies between product variants,'' in \emph{Intl Conference on Feature
  Interactions in Telecommunications and Software Systems VIII, ICFI'05, 28-30
  June 2005, Leicester, UK}.\hskip 1em plus 0.5em minus 0.4em\relax IOS Press,
  2005, pp. 198--216.

\bibitem{ArcegaFHC16}
L.~Arcega, J.~Font, {\O}.~Haugen, and C.~Cetina, ``Achieving knowledge
  evolution in dynamic software product lines,'' in \emph{23rd Intl Conf. on
  Software Analysis, Evolution, and Reengineering, {SANER} 2016, Suita, Osaka,
  Japan, March 14-18, 2016}.\hskip 1em plus 0.5em minus 0.4em\relax {IEEE}
  Comp. Soc., 2016, pp. 505--516.

\bibitem{ChenPYNZ14}
B.~Chen, X.~Peng, Y.~Yu, B.~Nuseibeh, and W.~Zhao, ``Self-adaptation through
  incremental generative model transformations at runtime,'' in \emph{36th Intl
  Conf on Softw. Eng., {ICSE} '14, Hyderabad, India - May 31 - June 07,
  2014}.\hskip 1em plus 0.5em minus 0.4em\relax {ACM}, 2014, pp. 676--687.

\bibitem{RamirezCMB10}
A.~J. Ramirez, B.~H.~C. Cheng, P.~K. McKinley, and B.~E. Beckmann,
  ``Automatically generating adaptive logic to balance non-functional tradeoffs
  during reconfiguration,'' in \emph{7th Intl Conf. on Autonomic Computing,
  {ICAC} 2010, Washington, DC, USA, June 7-11, 2010}.\hskip 1em plus 0.5em
  minus 0.4em\relax {ACM}, 2010, pp. 225--234.

\bibitem{Sousa_SEAMS_2017}
G.~Sousa, W.~Rudametkin, and L.~Duchien, ``Extending dynamic software product
  lines with temporal constraints,'' in \emph{12th Intl. Symposium on Softw.
  Eng. for Adaptive and Self-Managing Systems, {SEAMS '17}, Buenos Aires,
  Argentina, May 22-23, 2017}.\hskip 1em plus 0.5em minus 0.4em\relax
  Piscataway, NJ, USA: IEEE Press, 2017, pp. 129--139.

\end{thebibliography}
%
%
%

%

\begin{IEEEbiography}[{\includegraphics[width=1in,height=1.25in,clip,keepaspectratio]{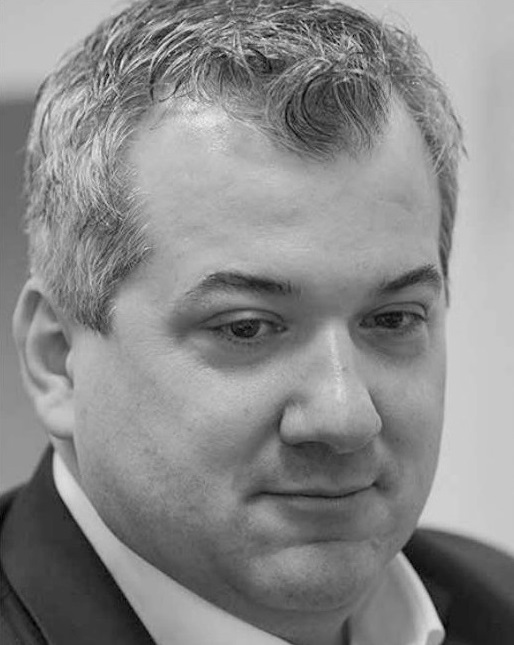}}]{Andreas Metzger} is senior academic councillor at the University of Duisburg-Essen and head of adaptive systems and big data applications at paluno, the Ruhr Institute for Software Technology. 
He is steering committee vice-chair of the European Technology Platform on Software, Services, Cloud and Data (NESSI) and deputy secretary general of the EU Big Data Value Association (BDVA). 
His background and research interests are software engineering for data-intensive and self-adaptive systems.
\end{IEEEbiography}
\vspace{-3em}

\begin{IEEEbiography}[{\includegraphics[width=1in,height=1.25in,clip,keepaspectratio]{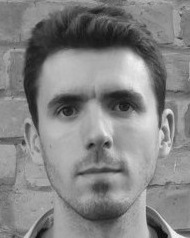}}]{Cl{\'{e}}ment Quinton} is associate professor in the Spirals group at University of Lille. His research activities focus on highly-configurable and distributed software systems operating in uncertain and evolving environments with the aim to devise formal models, methods and tools to cope with run-time evolution and adaptation of such systems.
\end{IEEEbiography}
\vspace{-3em}

\begin{IEEEbiography}[{\includegraphics[width=1in,height=1.25in,clip,keepaspectratio]{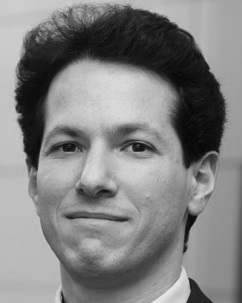}}]{Zolt{\'{a}}n {\'{A}}d{\'{a}}m Mann} is senior researcher at paluno, the Ruhr Institute for Software Technology of the University of Duisburg-Essen. 
He received his PhD in Computer Science from Budapest University of Technology and Economics, Hungary. 
His research interests include software engineering for self-adaptive systems and adaptive resource management in cloud computing.
\end{IEEEbiography}
\vspace{-3em}

\begin{IEEEbiography}[{\includegraphics[width=1in,height=1.25in,clip,keepaspectratio]{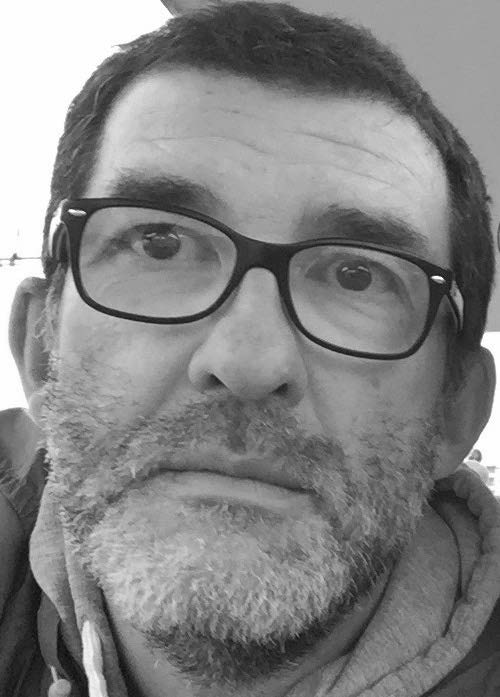}}]{Luciano Baresi} is full professor at the Politecnico di Milano.  Luciano was visiting professor at the University of Oregon (USA) and visiting researcher at the University of Paderborn (Germany). His research interests are in the broad area of software engineering and include formal approaches for modeling and specification languages, distributed systems, service-based applications and mobile, self-adaptive, and pervasive software systems. 
\end{IEEEbiography}
\vspace{-3em}

\begin{IEEEbiography}[{\includegraphics[width=1in,height=1.25in,clip,keepaspectratio]{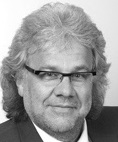}}]{Klaus Pohl} is full professor for software systems engineering at the University of Duisburg-Essen and director of paluno, the Ruhr Institute for Software Technology.
From 2005 to 2007 he was a founding director of lero, the Irish Software Engineering Research Centre. 
His research interests include requirements engineering, software product line engineering and engineering of self-adaptive systems.
\end{IEEEbiography}
\vspace{-3em}







\newpage
\section*{Supplemental Material}
\subsection*{Feature Models of the Exemplar Systems}
\label{sec:appendix}

The experiments presented in the paper build on four real-world exemplar systems and datasets.

\begin{itemize}
	\item CloudRM: A parametrized cloud resource management system (from~\cite{MannM17}).
	\item BerkeleyJ: The reconfigurable Berkeley database written in Java (from~\cite{Siegmund_ICSE_2012}).
	\item BerkeleyC: The reconfigurable Berkeley database written in C (from~\cite{Siegmund_ICSE_2012}).
	\item LLVM: The reconfigurable version of the LLVM compiler (from~\cite{Siegmund_ICSE_2012}).
\end{itemize}

As depicted in Table~\ref{table:datasetXX}, the systems differ with respect to the size of the adaptation space (\ie the number of feature combinations), the number of features and the depth of the feature model.

\begin{table}[h!]
 \centering
     \renewcommand{\arraystretch}{1.5}
\begin{tabular}{| c |  r | r | r |}
\hline
& size of  adap-& number of &  feature model\\
& tation space & features & depth \\
\hline
CloudRM 		&  344  &  63  &  3      \\
BerkeleyJ 		&  360  &  26  &  5        \\
LLVM 		&  1024  &  11  &  1      \\
BerkeleyC         &  2560  &  18  &  2     \\
\hline
\end
{tabular}
\caption{Systems and datasets used for the experiments.}
\label{table:datasetXX}
\end{table}

\begin{figure}[htbp]
\centering
\includegraphics[width=.9\columnwidth]{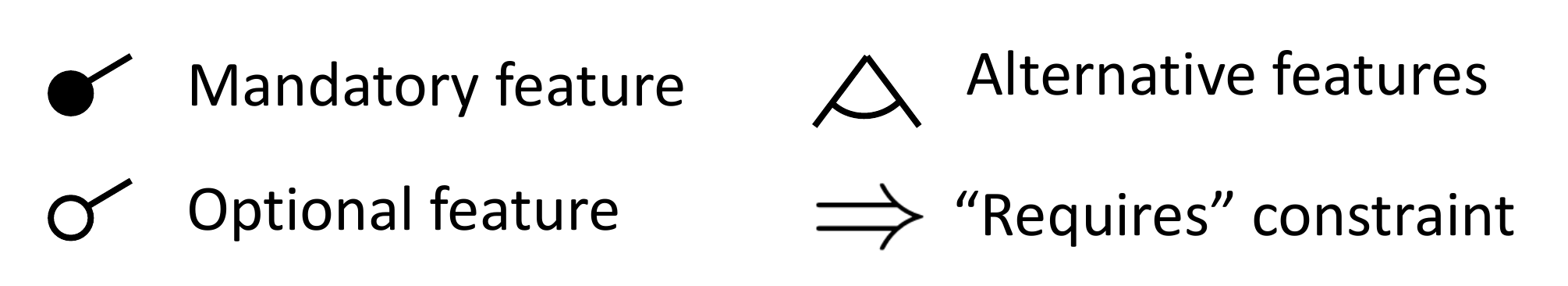}
\caption{Key to symbols for feature models}
\label{fig:key}
\end{figure}

The feature models of all four systems are provided on the following page.
Figure~\ref{fig:key} provides a key to the symbols used in the feature models.

\begin{figure*}[h!]
\centering
\includegraphics[width=2\columnwidth]{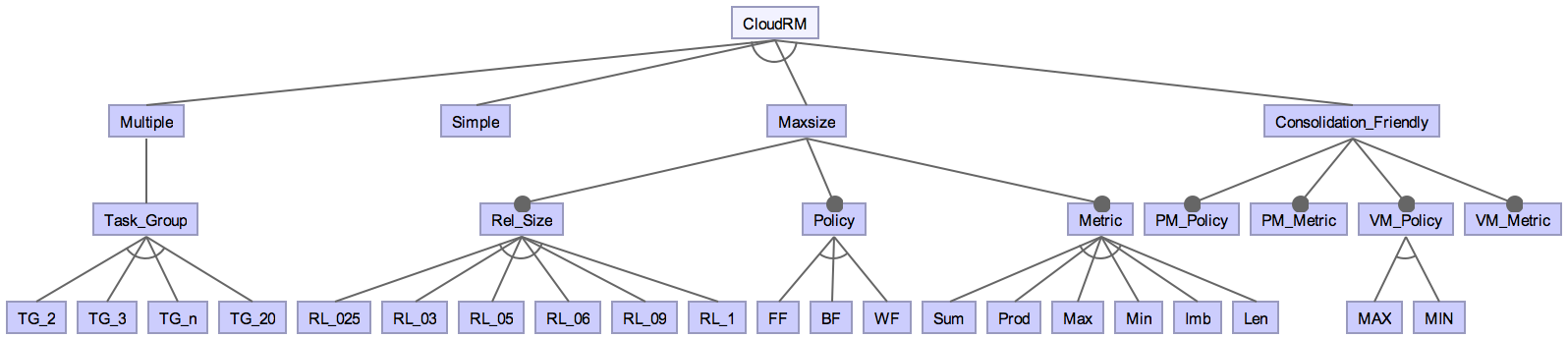}
\caption{Feature model of CloudRM}
\label{fig:Acloud}
\end{figure*}

Note that, due to space limitation, the entire \emph{CloudRM} feature model cannot fit in a readable format. 
We thus depict a reduced version of this feature model, where the sub-features of \mm{PM\_Policy}, \mm{PM\_Metric} and \mm{VM\_Metric} are not visible, and not all sub-features of \mm{Task\_Group} are shown. 
Instead, we describe below the missing features: 
\begin{itemize}
	\item \mm{PM\_Policy} has the alternative sub-features \mm{FF}, \mm{BF} and \mm{WF} (for First-Fit, Best-Fit and Worst-Fit respectively).
  \item \mm{PM\_Metric} and \mm{VM\_Metric} each have the alternative sub-features \mm{Sum}, \mm{Prod}, \mm{Max}, \mm{Min}, \mm{Imb} and \mm{Len} (similarly to the \mm{Metric} feature of the \mm{Maxsize} placement policy).
 \item \mm{Task\_Group} has 19 alternative sub-features \mm{TG\_$n$}, with $n = 2, \ldots, 20$.
\end{itemize}


\begin{figure*}[h!]
	\centering
		\includegraphics[width=1.5\columnwidth]{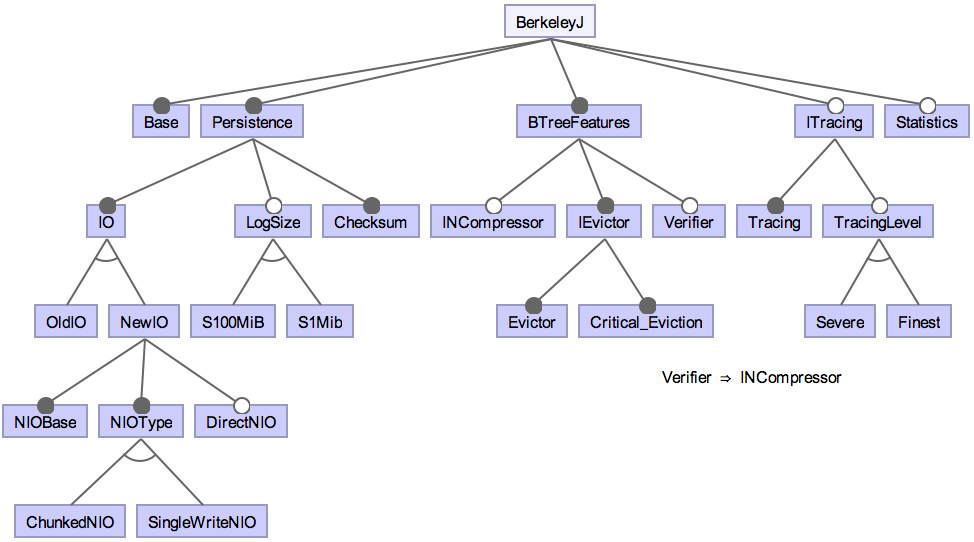}
	\caption{Feature model of BerkeleyJ}
	\label{fig:ABerkeleyJ}
\end{figure*}


\begin{figure*}[h!]
	\centering
		\includegraphics[width=1.8\columnwidth]{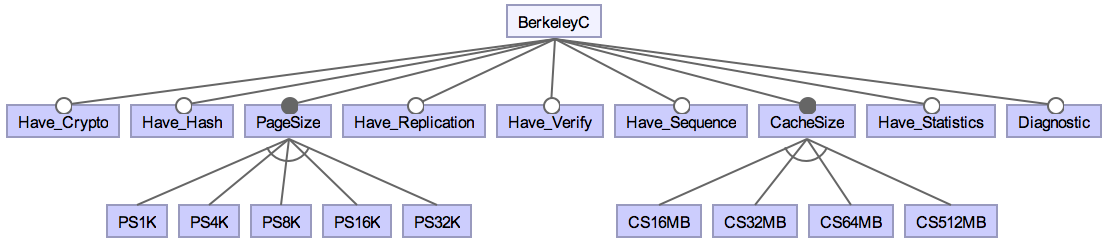}
	\caption{Feature model of BerkeleyC}
	\label{fig:ABerkeleyC}
\end{figure*}

~
\begin{figure*}[!h]
	\centering
		\includegraphics[width=1.8\columnwidth]{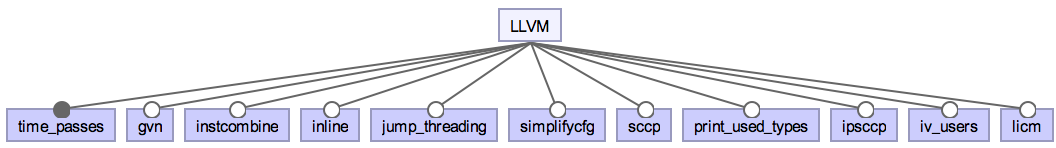}
	\caption{Feature model of LLVM}
	\label{fig:ALLVM}
\end{figure*}

\end{document}